\documentclass[a4paper,12pt]{article}%
\usepackage{amsmath}
\usepackage{amsfonts}
\usepackage{amssymb}
\usepackage{graphicx}%
\begin{document}

\title{The Modified J-Matrix Approach for Cluster Descriptions of Light Nuclei}
\author{F. Arickx$^{1}$, J. Broeckhove$^{1}$, A. Nesterov$^{2}$,\\V. Vasilevsky$^{2}$ and W. Vanroose$^{1}$ \\$^{1}$University of Antwerp, \\Group Computational Modeling and Programming, \\Antwerp, Belgium\\$^{2}$Bogolyubov Institute for Theoretical Physics, Kiev, Ukraine}
\date{}
\maketitle

\begin{abstract}
We present a fully microscopic three-cluster nuclear model for light nuclei on
the basis of a J-Matrix approach. We apply the Modified J-Matrix method on
$^{6}He$ and $^{6}Be$ for both scattering and reaction problems, analyse the
Modified J-Matrix calculation, and compare the results to experimental data.

\end{abstract}

\section{Introduction}

The J-Matrix method (JM) has proven very successful in microscopic nuclear
calculations, particularly the Modified J-Matrix approach
(MJM)\cite{2002PhRvL..88a0404V,kn:VA_PR}, also called the Algebraic Model in
some nuclear physics literature. Both collective \cite{kn:Vasil92} and cluster
descriptions \cite{kn:ITP+RUCA1,kn:ITP+RUCA2}, of light (p-shell) nuclei have
been studied with the MJM, using an oscillator basis.

The Harmonic Oscillator basis has always been very popular in nuclear physics.
For light nuclei, spherical oscillator states have often been used as a first
approximation to the single-particle orbital wave functions in the popular
nuclear shell-model. The nuclear many-body basis is then built up as a set of
Slater determinants of single-particle oscillator orbital states to take the
Pauli principle into account.

Many nuclear two-body potentials feature a superposition of Gaussian
components. Matrix elements of two-body operators for Slater determinants
reduce to a simple sum of two-body matrix elements, involving the
single-particle oscillator states. A Gaussian form of the operator then easily
leads to an analytical form for the matrix elements. This immediately shows
the computational advantage of using an oscillator state (or even a
superposition of oscillator states) for the single-particle wave functions in
a fully microscopic many-body nuclear model.

One of the features of the oscillator basis is the Jacobi (tridiagonal) form
of the matrix of the kinetic energy operator. This makes it a proper candidate
for considering the JM approach to solve the Schr\"{o}dinger equation
expressed in matrix form.

The nuclear many-body system is very complex though, and requires huge
superpositions of shell-model states to reproduce spectral properties over an
important energy range. Also, the nuclear system exhibits several modes when
the system is energetically excited, and possibly fragmented. There is an
interplay between collective modes, such as monopole and quadrupole
excitations, and cluster effects that are particularly pronounced when the
nucleus disintegrates. One therefore often introduces very specific
antisymmetrized forms for the wave function with a specific configuration of
single-particle orbitals, that feature some specific collective behavior that
one wants to study. In this way the Hilbert space is limited to a single (or a
few coupled) nuclear model state(s) in which the collective coordinates are
the only remaining dynamical coordinates. The dimensions of the
Schr\"{o}dinger equation are then strongly reduced.

Although the nuclear two-body interaction has a short range, the effective
interaction as a function of the collective coordinates often displays a long
range. If the collective behavior is described as a superposition of
oscillator states, this is usually also reflected in the matrix elements of
the nuclear potential, which display a slow decrease for increasing oscillator
excitation. This clearly limits the applicability of the standard JM method,
as too large energy matrices have to be considered in the internal,
non-asymptotic, region. Indeed, the main computational cost for nuclear JM
calculations often lies in the construction of the matrix elements of highly
excited states. To account for this limitation the MJM approach was developed
\cite{2002PhRvL..88a0404V,kn:VA_PR}, introducing a semi-classical
approximation for the matrix elements in the highly excited internal region.
This leads to a modification of the standard three-term recursion relation for
the far interaction and asymptotic region by including (semiclassical)
potential contributions. A drastic reduction of the matching position for
boundary condition is obtained, resulting in a remaining matrix equation of
manageable dimensions.

The Coulomb contribution to the nuclear potential, which is known to have a
long range, can be handled in the same way.

In the following sections we will elaborate on the cluster description of
nuclei as an important collective description for the lightest nuclei, in
which the internuclear distances then become the dynamical (collective)
coordinates. We will link this model to the MJM approach to solve the
Schr\"{o}dinger equation, and present applications on 6-particle nuclear
systems where three-cluster effects are important. We will compare the
theoretical results to the experimental ones to test the validity of the approach.

\section{JM Cluster models for light nuclei}

\label{sect:JMclusterModel}

Two- and three-cluster configurations carry an important part of the
low-energy physics in light nuclei, and directly relate to scattering and
reaction experiments in which smaller fragments are used to study compound properties.

In this section we discuss the three-cluster description for light nuclei.
Where appropriate we briefly present some two-cluster properties, as the
three-cluster approach is essentially a generalization hereof.

\begin{figure}
[ptb]
\begin{center}
\includegraphics[
trim=0.000000in 0.099015in 0.000000in 0.099442in,
height=4.2043cm,
width=8.4416cm
]%
{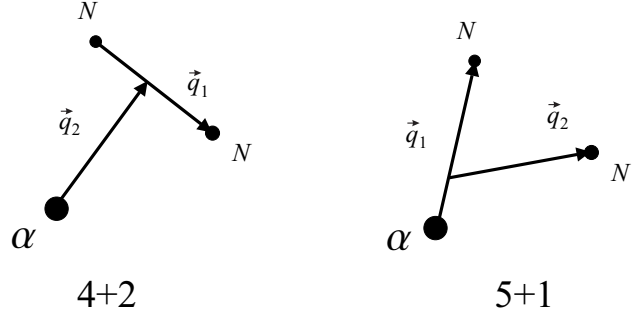}%
\caption{Two configurations of Jacobi coordinates for the three-cluster system
$\alpha+N+N$.}%
\label{fig:figure1}%
\end{center}
\end{figure}

The many-particle wave functions for a three-cluster system of $A$ nucleons
($A=A_{1}+A_{2}+A_{3}$) can be written, using the anti-symmetrization operator
$\mathcal{A}$, as follows%

\begin{equation}
\Psi\left(  \mathbf{q}_{1},..,\mathbf{q}_{A-1}\right)  =\mathcal{A}\left[
\Psi_{1}\left(  A_{1}\right)  \ \Psi_{2}\left(  A_{2}\right)  \ \Psi
_{3}\left(  A_{3}\right)  \ \Psi_{R}\left(  R\right)  \right]
\label{eq:AntisymClustState}%
\end{equation}
where the centre of mass of the $A$-nucleon system has been eliminated by the
use of Jacobi coordinates $\mathbf{q}_{i}$ so that only internal dynamics are
described. The cluster wave functions $\Psi_{i}\left(  A_{i}\right)  $
\begin{equation}
\Psi_{i}\left(  A_{i}\right)  =\Psi_{i}\left(  \mathbf{q}_{1}^{\left(
i\right)  },..,\mathbf{q}_{A_{i}-1}^{\left(  i\right)  }\right)
\quad(i=1,2,3) \label{eq:ClustState}%
\end{equation}
represent the internal structure of the $i$-th cluster, centered around its
centre of mass $\mathbf{R}_{i}$. To limit the computational complexity of the
problem, these cluster functions are fixed and they are Slater determinants of
harmonic oscillator ($0s$)-states, corresponding to the groundstate
shell-model configuration of the cluster ($A_{i}\leq4$ for all $i$). The
$\Psi_{R}\left(  R\right)  $ wave function
\begin{equation}
\Psi_{R}\left(  R\right)  =\Psi_{R}\left(  \mathbf{q}_{1}^{\left(  R\right)
},\mathbf{q}_{2}^{\left(  R\right)  }\right)  =\Psi_{R}\left(  \mathbf{q}%
_{1},\mathbf{q}_{2}\right)  \label{eq:RelMotionState}%
\end{equation}
represents the relative motion of the three clusters with respect to one
another, and $\mathbf{q}_{1}$ and $\mathbf{q}_{2}$ represent Jacobi
coordinates. In figure \ref{fig:figure1} we indicate an enumeration of
possible Jacobi coordinates and their relation to the component clusters.%

The state (\ref{eq:RelMotionState}) is not limited to any particular type of
orbital; on the contrary we will use a complete basis of harmonic oscillator
states for the relative motion degrees of freedom. Thus the full $A$-particle
state cannot be expressed as a single Slater determinant of single particle orbitals.

An important approximation, known as the \textquotedblleft
Folding\textquotedblright\ model, is obtained by breaking the Pauli principle
between the individual clusters, but retaining a proper quantum-mechanical
description of the clusters, which is described by the wave function
\begin{equation}
\Psi_{F}\left(  \mathbf{q}_{1},..,\mathbf{q}_{A-1}\right)  =\Psi_{1}\left(
A_{1}\right)  \ \Psi_{2}\left(  A_{2}\right)  \ \Psi_{3}\left(  A_{3}\right)
\ \Psi_{R}\left(  R\right)
\end{equation}
Because each cluster wave function is antisymmetric (they are Slater
determinants) one is indeed neglecting the inter-cluster anti-symmetrization
only. The Folding Model has the advantage of preserving the identities of the
clusters and, if the intra-cluster structure is kept \textquotedblleft
frozen\textquotedblright, it reduces the many-particle problem to that of the
relative motion of the clusters.

For a two-cluster description the formulae simplify to:%
\begin{equation}
\Psi\left(  \mathbf{q}_{1},..,\mathbf{q}_{A-1}\right)  =\mathcal{A}\left[
\Psi_{1}\left(  A_{1}\right)  \ \Psi_{2}\left(  A_{2}\right)  \ \Psi
_{R}\left(  R\right)  \right]  \label{eq:AntisymClustState2}%
\end{equation}
with%
\begin{equation}
\Psi_{R}\left(  R\right)  =\Psi_{R}\left(  \mathbf{q}_{0}^{\left(  R\right)
}\right)  =\Psi_{R}\left(  \mathbf{q}_{0}\right)  \label{eq:RelMotionState2}%
\end{equation}

The folding approximation will be the natural choice for calculating the
asymptotic behavior of the cluster-system, i.e. the disintegration of the
system in the two or three non-interacting individual clusters. This amounts
to the situation that all clusters are a sufficient distance apart and
inter-cluster antisymmetrization has a negligible effect.

Because the cluster states are fixed and built up of ($0s$)-orbitals, the
problem of labeling the basis states with quantum numbers relates to the
inter-cluster wave function only. This holds true whether one uses the full
anti-symmetrization or the folding approximation. In a two-cluster case, the
set of quantum numbers describing inter-cluster motion is unambiguously
defined, and is obtained from the reduction of the symmetry group $U(3)\supset
O(3)$ of the one-dimensional oscillator. This reduction provides the quantum
numbers $n$ for the radial excitation, and $L,M$ for the angular momentum of
the two-cluster system.

In a three-cluster case, several schemes can be used to classify the
inter-cluster wave function in the oscillator representation. In
\cite{kn:Vasil96,kn:Vasil97} three distinct but equivalent schemes were
considered. One of these used the quantum numbers provided by the
Hyperspherical Harmonics (HH) method (see for instance
\cite{kn:Simon68E,kn:Fabr93,kn:Zhuk93}). This is the classification that we
will adopt. Even within this particular scheme there are several ways to
classify the basis states. We shall restrict ourselves to the so-called
Zernike-Brinkman basis \cite{kn:ZB}. This corresponds to the following
reduction of the unitary group $U(6)$, the symmetry group of the
three-particle oscillator Hamiltonian,
\begin{equation}
U(6)\supset O(6)\supset O(3)\otimes O(3)\supset O(3) \label{eq:GroupReduc}%
\end{equation}
This reduction provides the quantum numbers $K$, the hypermomentum, $n$, the
hyperradial excitation, $l_{1}$, the angular momentum connected with the first
Jacobi vector, $l_{2}$, the angular momentum connected with the second Jacobi
vector, and $L$ and $M$ the total angular momentum obtained from coupling the
partial angular momenta $l_{1}$, $l_{2}$. Collectively these quantum numbers
will be denoted by $\nu$, i.e. $\nu=\{n,K,(l_{1}l_{2})LM\}$ in the remainder
of the text.

There are a number of relations and constraints on these quantum numbers:

\begin{itemize}
\item the total angular momentum is the vector sum of the partial angular
momenta $\mathbf{l}_{1}$and $\mathbf{l}_{2}$, i.e. $\mathbf{L}=$
$\mathbf{l}_{1}+$ $\mathbf{l}_{2}$ or $\left|  l_{1}-l_{2}\right|  \leq L\leq
l_{1}+l_{2}$.

\item by fixing the values of $l_{1}$ and $l_{2}$, we impose restrictions on
the hypermomentum $K=l_{1}+l_{2},~l_{1}+l_{2}+2,~l_{1}+l_{2}+4,\ldots$ This
condition implies that for certain values of hypermomentum $K$ the sum of
partial angular momenta $l_{1}+l_{2}$ cannot exceed $K$.

\item the partial angular momenta $l_{1}$ and $l_{2}$ define the parity of the
three-cluster state by the relation $\pi=\left(  -1\right)  ^{l_{1}+l_{2}} $.

\item for the ``normal'' parity states $\pi=\left(  -1\right)  ^{L}$ the
minimal value of hypermomentum is $K_{\min}=L$, whereas $K_{\min}=L+1$ for the
so-called ``abnormal'' parity states $\pi=\left(  -1\right)  ^{L+1}$.

\item oscillator shells with $N$ quanta are characterized by the constraint
$N=2n\;+K$.
\end{itemize}

Thus for a given hyperangular and rotational configuration the quantumnumber
$n$ ladders the oscillator shells of increasing oscillator energy.

\subsection{Asymptotic solutions in coordinate representation}

\label{sect:AsympSolCoordRep}

The MJM method for solving the Schr\"{o}dinger equation for quantum scattering
systems is based on a matrix representation of the Schr\"{o}dinger equation in
terms of a square integrable basis. In a nuclear context the Harmonic
Oscillator basis is appropriate. The boundary conditions are ultimately
formulated in terms of the asymptotic behavior of the expansion coefficients
of the wave function.

In the case of three-cluster calculations, one needs to determine the
asymptotic behavior of the wave function. Consider an expansion of the
relative wave function
\begin{equation}
\Psi_{R}\left(  \mathbf{q}_{1},\mathbf{q}_{2}\right)  =\sum_{\nu}c_{\nu}%
\Psi_{\nu}\left(  \mathbf{q}_{1},\mathbf{q}_{2}\right)
\label{eq:RelWaveFuncAsympt}%
\end{equation}
with $\nu=\{n,K,(l_{1}l_{2})LM\}$ and $\left\{  \Psi_{\nu}\right\}  $ a
complete basis of six-dimensional oscillator states. It covers all possible
types of relative motion between the three clusters.

To obtain the asymptotic behavior of the three-cluster system the folding
approximation is used. The assumption that antisymmetrization effects between
clusters are absent in the asymptotic region is a natural one. It simplifies
the many-body dynamics effectively to that of a three-particle system, as the
cluster descriptions are frozen. The relative motion problem of the three
clusters in the absence of a potential (i.e. considering the JM reference
Hamiltonian only) can then be explicitly solved in the HH method (see for
instance \cite{kn:Simon68E,kn:Fabr93,kn:Zhuk93}). It involves the
transformation of the Jacobi coordinates $\mathbf{q}_{1}$ and $\mathbf{q}_{2}
$\ to the hyperradius $\rho$\ and a set of hyperangles $\Omega$. The
hyperspherical coordinates $\rho$\ and $\Omega$ describe the geometry of the
three-particle system in the same way that the spherical coordinates describe
the geometry of two-particle systems. The inter-cluster wave function in
coordinate representation is expanded in HH's $H_{K}^{\nu_{0}}\left(
\Omega\right)  $ where $\nu_{0}$ has been chosen as a shorthand for
$(l_{1}l_{2})LM$, and which are the generalization of the spherical harmonics
$Y_{LM}\left(  \theta,\varphi\right)  $.

In the absence of the Coulomb interaction this leads to a set of equations for
the hyperradial asymptotic solutions, with the kinetic energy operator as the
JM reference Hamiltonian
\begin{equation}
\left\{  -\frac{\hbar^{2}}{2m}\left[  \frac{d^{2}}{d\rho^{2}}+\frac{5}{\rho
}\frac{d}{d\rho}-\frac{K\left(  K+4\right)  }{\rho^{2}}\right]  -E\right\}
R_{K,\nu_{0}}\left(  \rho\right)  =0
\end{equation}

The solutions can be obtained analytically and are represented by a pair of
H\"{a}nkel functions for the ingoing and outgoing solutions:
\begin{equation}
R_{K,\nu_{0}}^{\left(  \pm\right)  }\left(  \rho\right)  =\left\{
\begin{array}
[c]{l}%
H_{K+2}^{\left(  1\right)  }\left(  k\rho\right)  /\rho^{2}\\
H_{K+2}^{\left(  2\right)  }\left(  k\rho\right)  /\rho^{2}%
\end{array}
\right\}
\end{equation}
where
\[
k=\sqrt{\frac{2mE}{\hbar^{2}}}%
\]
One notices that these asymptotic solutions are independent of all quantum
numbers $\nu_{0}$, and are determined by the value of hypermomentum $K$ only.
In particular, different $K$-channels are uncoupled.

When charged clusters are considered the asymptotic reference Hamiltonian
should consist of the kinetic energy and the Coulomb interaction:%

\begin{equation}
\left\{  -\frac{\hbar^{2}}{2m}\left[  \frac{d^{2}}{d\rho^{2}}+\frac{5}{\rho
}\frac{d}{d\rho}-\frac{\left\Vert \mathcal{K}\right\Vert }{\rho^{2}}\right]
+\frac{\left\Vert Z_{eff}\right\Vert }{\rho}-E\right\}  \left\Vert
\mathcal{R}\left(  \rho\right)  \right\Vert =0
\end{equation}
The matrix $\left\Vert \mathcal{K}\right\Vert $ is diagonal with matrix
elements $K\left(  K+4\right)  $, and $\left\Vert Z_{eff}\right\Vert $, the
\textquotedblleft effective charge\textquotedblright, is off-diagonal in $K$
and $(l_{1}l_{2})$. The solution matrix $\left\Vert \mathcal{R}\left(
\rho\right)  \right\Vert $ reflects the coupling of $K$-channels. A standard
approximation for solving these equations is to decouple the $K$-channels, by
assuming that the off-diagonal matrix-elements of $\left\Vert Z_{eff}%
\right\Vert $ are sufficiently small:
\begin{equation}
\left\{  -\frac{\hbar^{2}}{2m}\left[  \frac{d^{2}}{d\rho^{2}}+\frac{5}{\rho
}\frac{d}{d\rho}-\frac{K\left(  K+4\right)  }{\rho^{2}}\right]  +\frac
{Z_{eff}}{\rho}-E\right\}  R_{K,\nu_{0}}\left(  \rho\right)  =0
\end{equation}
The constants $Z_{eff}$ depends on $K$ and $\nu_{0}$ and all parameters of the
many-body system under consideration. We will restrict ourselves to this
decoupling approximation, but it is to be understood that its validity has to
be checked for any specific three-cluster system.

The asymptotic solutions then become
\begin{equation}
R_{K,\nu_{0}}^{\left(  \pm\right)  }\left(  \rho\right)  =\frac{1}{\sqrt{k}%
}W_{\pm i\eta,K+2}\left(  \pm2ik\rho\right)  /\rho^{\frac{5}{2}}%
\end{equation}
where $W$ is the Whittaker function and $\eta$\ is the well-known Sommerfeld
parameter
\begin{equation}
\eta=\frac{m}{2\hbar^{2}}\frac{Z_{eff}}{k}%
\end{equation}

As $\eta$\ is a function of $K,$ $l_{1}$ and $l_{2}$ through the parameter
$Z_{eff}$, the asymptotic solutions will now be dependent on $K$ and $\nu_{0}
$.

A two-cluster description leads to the well-known one-dimensional radial
equation, with asymptotic solutions%
\[
R_{L}^{\left(  \pm\right)  }\left(  \rho\right)  =\frac{1}{\sqrt{k}}W_{\pm
i\eta,L+\frac{1}{2}}\left(  \pm2ik\rho\right)  /\rho
\]
where $\rho$ now is the inter-cluster radial distance and $\eta$ is the
Sommerfeld parameter with $Z_{eff}=Z_{1}Z_{2}e^{2}\sqrt{\frac{A_{1}A_{2}%
}{A_{1}+A_{2}}}$

One can easily combine the two- and three cluster asymptotics in a single
representation as%
\begin{equation}
R_{\mathcal{L}}^{\left(  \pm\right)  }\left(  \rho\right)  =\frac{1}{\sqrt{k}%
}W_{\pm i\eta,\lambda}\left(  \pm2ik\rho\right)  /\rho^{\frac{\sigma-1}{2}}%
\end{equation}
where the parameters $\mathcal{L}$, $\lambda$, $\sigma$\ and $\eta$\emph{,}
differ for the two- and three-cluster channels, and are summarized in Table
\ref{tab:HH state}.%

\begin{table}[tbp] \centering
\caption{Two- and three-cluster asymptotic solution parameters}
\begin{tabular}
[c]{|c|c|c|c|c|}\hline
& $\mathcal{L}$ & $\sigma$ & $\lambda$ & $\eta$\\\hline
two-$\text{cl}$. channel & $L$ & 3 & $L+\frac{1}{2}$ & $\frac{Z_{1}Z_{2}e^{2}%
}{k}\frac{m}{2\hbar^{2}}\sqrt{\frac{A_{1}A_{2}}{A_{1}+A_{2}}}$\\\hline
three-$\text{cl. channel}$ & $K$ & 6 & $K+2$ & $\frac{Z_{eff}}{k}\frac
{m}{2\hbar^{2}}$\\\hline
\end{tabular}
\label{tab:TwoThreeClAsymps}
\end{table}%

\subsection{Asymptotic solutions in oscillator representation}

The JM relies on an expansion in terms of oscillator functions, and the
asymptotic behavior of the corresponding expansion coefficients $c_{\nu}$. We
will restrict ourselves to the three-cluster situation; the two-cluster
results can be readily obtained from \cite{kn:VA_PR} and the results of the
section \ref{sect:AsympSolCoordRep}.

It was conjectured (see for instance \cite{kn:VA_PR}) that for very large
values of the oscillator quantum number $n$ the expansion coefficients for
physically relevant wave-functions behave like
\begin{equation}
c_{n}=\left\langle n|\psi\right\rangle \simeq\sqrt{2}\rho_{n}^{2}\psi
(b\rho_{n}) \label{eq:AsymptoticForCnGeneral}%
\end{equation}
with $b$ the oscillator parameter of the basis, $\rho_{n}=\sqrt{4n+2K+6}$ the
classical turning point, and $\psi$ the hyperradial wave function.

In the case of neutral clusters this leads after substitution of the
hyperradial asymptotic solutions to the following expansion coefficients
$c_{n}^{\left(  \pm\right)  }$
\begin{equation}
c_{n}^{\left(  \pm\right)  K}\simeq\sqrt{2}\left\{
\begin{array}
[c]{l}%
H_{K+2}^{\left(  1\right)  }\left(  kb\rho_{n}\right) \\
H_{K+2}^{\left(  2\right)  }\left(  kb\rho_{n}\right)
\end{array}
\right\}  \label{eq:AsymCoefFreePart}%
\end{equation}

This result can be obtained in an alternative way \cite{kn:4n,kn:AM_12C} by
representing the Schr\"{o}dinger equation, with the kinetic energy operator
$\hat{T}$ as the reference Hamiltonian to describe the asymptotic situation,
in a (hyperradial) oscillator representation
\begin{equation}
\sum_{m=0}^{\infty}\left\langle n,\left(  K,\nu_{0}\right)  \left\vert \hat
{T}-E\right\vert m,\left(  K,\nu_{0}\right)  \right\rangle c_{m}^{K,\nu_{0}}=0
\end{equation}
This matrix equation is of a three-diagonal form because of the properties of
$\hat{T}$ and the oscillator basis. Solving for the expansion coefficients
$c_{n}^{K,\nu_{0}}$ leads to a three-term recurrence relation
\begin{equation}
T_{n,n-1}^{K,\nu_{0}}c_{n-1}^{K,\nu_{0}}+\left(  T_{n,n}^{K,\nu_{0}}-E\right)
c_{n}^{K,\nu_{0}}+T_{n,n+1}^{K,\nu_{0}}c_{n+1}^{K,\nu_{0}}=0
\end{equation}
where
\begin{equation}
T_{n,m}^{K,\nu_{0}}=\left\langle n,\left(  K,\nu_{0}\right)  \left\vert
\hat{T}\right\vert m,\left(  K,\nu_{0}\right)  \right\rangle
\end{equation}
The asymptotic solutions (i.e. for high $n$) of this recurrence relation are
then precisely given by (\ref{eq:AsymCoefFreePart}).

When the Coulomb interaction is present we again apply
(\ref{eq:AsymptoticForCnGeneral}) to obtain
\begin{equation}
c_{n}^{\left(  \pm\right)  K}\simeq\sqrt{2}\left\{
\begin{array}
[c]{l}%
W_{i\eta,\mu}\left(  2ikb\rho_{n}\right)  /\sqrt{\rho_{n}}\\
W_{-i\eta,\mu}\left(  -2ikb\rho_{n}\right)  /\sqrt{\rho_{n}}%
\end{array}
\right\}  \label{eq:AsymCoefCoulPart}%
\end{equation}
In this case the oscillator representation of the Schr\"{o}dinger equation is
no longer of a tridiagonal form, and cannot be solved analytically for the
asymptotic solutions to corroborate this result.

It should be noted that the above elaborations are valid for relatively small
values of momentum $k$ to which correspond sufficiently large values of the
discrete hyperradius $\rho_{n}$, which defines the asymptotic region. So for
any value of $k$ one can determine a minimum value for $n$ to be safely in the
asymptotic region.

As we consider an asymptotic decoupling in the $\left(  K,\nu_{0}\right)  $
quantumnumbers, one will deal with asymptotic channels characterized by the
$\left(  K,\nu_{0}\right)  $ values. So only in the internal (or interaction)
region will states with different $K$ and $\nu_{0}$ be coupled by the
short-range nuclear potential and the Coulomb potential. The three-cluster
system can therefore be described by a coupled-channels approach, where the
individual channels are characterized by a single $K$-value, and we will
henceforth refer to these channels as ``$K$-channels''.

\subsection{Multi-channel JM equations}

In the current many-channel description of the JM formulation for
three-cluster systems, the channels will be characterized by the a specific
value of the set of quantumnumbers $K,\nu_{0}$, whereas the relative motion of
clusters within the channel is connected to the oscillator index $n$. We will
use $K$ henceforth as an aggregate index for individual channels, and assume
it represents all $K,\nu_{0}$ quantum numbers.

The Schr\"{o}dinger equation can be cast in a matrix equation of the form
\begin{equation}
\sum_{K^{\prime},m}\left\langle n,K\left\vert \hat{H}-E\right\vert
m,K^{\prime}\right\rangle \ c_{m}^{K^{\prime}}=0 \label{eq:CouplChanEqs1}%
\end{equation}

As we will consider an $S$-matrix formulation of the scattering problem, the
expansion coefficients are rewritten as
\begin{equation}
c_{n}^{K}=c_{n}^{(0)K}+\delta_{K_{i}K}c_{n}^{(-)K}-S_{K_{i}K}c_{n}^{(+)K}
\label{eq:AsymptExpOfCn}%
\end{equation}
where, for each channel $K$, the $c_{n}^{(0)K}$\ are the so-called residual
coefficients, the $c_{n}^{(\pm)K}$ are the incoming and outgoing asymptotic
coefficients. The matrix element $S_{K_{i}K}$ describes the coupling between
the current channel $K$ and the entrance channel $K_{i}$.

As shown in \cite{kn:Heller1,kn:Yamani,kn:SmirnovE}, the $c_{n}^{(\pm)K}$
satisfy the following system of equations%
\begin{equation}
\sum_{m=0}^{\infty}\left\langle n,K\left\vert \hat{H}_{0}-E\right\vert
m,K\right\rangle \ c_{m}^{(\pm)K}=\beta_{0}^{(\pm)K}\ \delta_{n,0}
\label{eq:AsymptEqnsKchannels}%
\end{equation}
$\hat{H}_{0}$ being the asymptotic reference Hamiltonian, consisting of the
kinetic energy operator for uncharged clusters, and the kinetic energy
operator plus Coulomb interaction for charged clusters. The right-hand side
for the equation for $c_{m}^{(-)K}$ features $\beta_{0}^{(-)K}$ which is a
regularization factor to account for the irregular behavior of the
$c_{0}^{(-)K}$. This factor allows one to solve (\ref{eq:AsymptEqnsKchannels})
for all values of $n$. The value of $\beta_{0}^{(-)K}$ can be obtained for
both reference Hamiltonians (i.e. with or without Coulomb). We introduced
$\beta_{0}^{(+)K}=0$ to keep the equations in a homogeneous form.

The set of equations (\ref{eq:AsymptEqnsKchannels})\ for the asymptotic
coefficients can then be solved numerically to different degrees of
approximation depending on the requested precision. The $c_{n}^{\left(
\pm\right)  K\text{ }}$have the desired asymptotic behavior (cfr eqs
\ref{eq:AsymCoefFreePart} and \ref{eq:AsymCoefCoulPart}).

Substitution of (\ref{eq:AsymptExpOfCn}) in the equations
(\ref{eq:CouplChanEqs1}) then leads to the following system of dynamical
equations for the many-channel system:%
\begin{equation}
\sum_{K^{\prime},m}\left\langle n,K\left\vert \hat{H}-E\right\vert
m,K^{\prime}\right\rangle c_{m}^{\left(  0\right)  K^{\prime}}-\sum
_{K^{\prime}}S_{K_{i}K^{\prime}}V_{n}^{\left(  +\right)  KK^{\prime}}%
=-\beta_{0}^{(-)K}\ \delta_{n,0}\delta_{K_{i}K}-V_{n}^{\left(  -\right)
KK_{i}} \label{eq:CouplChanEqs2}%
\end{equation}
where the coefficients $V_{n}^{\left(  \pm\right)  KK^{\prime}}$, defined in
\cite{kn:VA_PR}, are given by
\begin{equation}
V_{n}^{(\pm)KK^{\prime}}=\sum_{m=0}^{\infty}\left\langle n,K\left\vert \hat
{V}\right\vert m,K^{\prime}\right\rangle \ c_{m}^{(\pm)K^{\prime}}%
\end{equation}

This system of equations should be solved for both the residual coefficients
$c_{n}^{\left(  0\right)  K}$ and the $S$-matrix elements $S_{K^{\prime}K}$.

To obtain an appropriate approximation to the exact solution of
(\ref{eq:CouplChanEqs2}), we consider an internal region corresponding to
$n<N$ and an asymptotic region with $n\geq N$. The choice of $N$ is such that
one can expect the residual expansion coefficients $\left\{  c_{n}%
^{(0)K}\right\}  $ to be sufficiently small in the asymptotic region. Under
these assumptions (\ref{eq:CouplChanEqs2}) reduces to the following set of
$N+1$ equations ($n=0..N$):
\begin{equation}
\sum_{K^{\prime},m<N}\left\langle n,K\left\vert \hat{H}-E\right\vert
m,K^{\prime}\right\rangle c_{m}^{(0)K^{\prime}}-\sum_{K^{\prime}}%
S_{K_{i}K^{\prime}}V_{n}^{\left(  +\right)  KK^{\prime}}=-\beta_{0}%
^{(-)K}\ \delta_{n,0}\delta_{K_{i}K}-V_{n}^{\left(  -\right)  KK_{i}}%
\end{equation}

The total number of equations for an entrance channel $K_{i}$ amounts to
$N_{ch}\ast\left(  N+1\right)  $, and solving the set of equations by
traditional numerical linear algebra leads to $N_{ch}\ast N$ residual
coefficients $\left\{  c_{n}^{\left(  0\right)  K}\text{; }K=K_{\min}%
..K_{\max}\text{; }n=0..N-1\right\}  $ and $N_{ch}$ $S$-matrix elements
$\left\{  S_{K_{i}K}\text{; }K=K_{\min}..K_{\max}\right\}  $. The set of
equations has to be solved for all $N_{ch}$ entrance channels labelled by
$K_{i}$.

\subsection{Matrix elements and the Generating Function Method}

We only present the general principles for calculating matrix elements in a
three-cluster basis. We will do so by considering the Generating Function
Method. The two main quantities of interest to solve the JM equations are: the
overlap matrix, and the Hamiltonian matrix. The former is of importance
because of the proper normalization of the basis states. The latter is
decomposed into the kinetic energy operator, the matrix elements of which are
obtained mainly by group-theoretical considerations, the potential energy
operator, which in our case will be chosen to be a local, semi-realistic,
two-body interaction based on a superposition of Gaussians, and the Coulomb interaction.

The basic principle of generating functions is well-known from mathematical
physics. A generating function or generator state depends on a parameter,
referred to as the generating coordinate, in such a way that an expansion with
respect to that parameter yields basis states as expansion terms. Let us
explain the principle using a familiar example: the single-particle translated
Gaussian wave functions, which are appropriate for two-cluster descriptions%

\begin{equation}
\phi(\mathbf{r}\left\vert \mathbf{R}\right.  )=\exp\left\{  -\frac{1}%
{2}\mathbf{r}^{2}+\sqrt{2}\ \mathbf{R}\cdot\mathbf{r}-\frac{1}{2}%
\mathbf{R}^{2}\right\}  \label{eq:sOrbital}%
\end{equation}
with the translation parameter $\mathbf{R}$ acting as generator coordinate.
The choice of parametrization of the generator coordinate influences the
quantum numbers of the individual basis states that are generated. In a
Cartesian parametrization $\mathbf{R}=(R_{x},R_{y},R_{z})$ one generates the
familiar Cartesian $\phi_{n_{x}}(R_{x})\phi_{n_{y}}(R_{y})\phi_{n_{z}}(R_{z})$
oscillator states. With a radial parametrization $\mathbf{R}=R\mathbf{\check
{R}}$ (where the inverted hat stands for a unit vector) the expansion yields
\begin{equation}
\phi(\mathbf{r}\left\vert \mathbf{R}\right.  )=\sum_{n,l,m}\mathcal{N}%
_{nl}R^{2n+l}Y_{lm}(\mathbf{\check{R})}\phi_{nlm}(\mathbf{r})
\label{eq:sOrbitalRad}%
\end{equation}
An underlying mathematical connection exists between such expansions, group
representation theory and coherent state analysis
\cite{perelom72,kn:perelomov}. We exploit the generating function principle to
facilitate the computation of matrix elements. The matrix element of any
operator between generating states is a function of the generating coordinates
on the left and right
\begin{equation}
X(\,\mathbf{R},\mathbf{R}^{\prime})=\left\langle \phi\left(  \mathbf{r}%
\left\vert \mathbf{R}\right.  \right)  \left\vert \mathbf{\hat{X}}\right\vert
\phi\left(  \mathbf{r}\left\vert \mathbf{R}^{\prime}\right.  \right)
\right\rangle \label{eq:GFMEgeneral}%
\end{equation}
Expansion of this function will yield the matrix elements between the basis
states. They can be identified in the expansion by the appropriate dependence
on the generator coordinates
\begin{equation}
X(\,\mathbf{R},\mathbf{R}^{\prime})=\sum_{nlm}\sum_{n^{\prime}l^{\prime
}m^{\prime}}\mathcal{N}_{nl}\mathcal{N}_{n^{\prime}l^{\prime}}R^{2n+l}%
R^{\prime(2n^{\prime}+l^{\prime})}Y_{lm}^{\ast}(\mathbf{\check{R}%
)}Y_{l^{\prime}m^{\prime}}(\mathbf{\check{R}}^{^{\prime}}\mathbf{)}%
\left\langle \phi_{nlm}(\mathbf{r})\left\vert \mathbf{\hat{X}}\right\vert
\phi_{n^{\prime}l^{\prime}m^{\prime}}(\mathbf{r})\right\rangle
\label{eq:GFMEExpand}%
\end{equation}

For the three-cluster basis we consider the following generating function for
inter-cluster basis functions (in what follows we shall use small $\mathbf{q}$
for the Jacobi vectors and capital $\mathbf{Q}$ for the corresponding
generating coordinates)%

\begin{equation}
\Psi\left(  \mathbf{q_{1},q_{2}}\left\vert \mathbf{Q_{1},Q_{2}}\right.
\right)  =\exp\left\{  -\frac{1}{2}\left(  \mathbf{q}_{1}^{2}+\mathbf{q_{2}%
^{2}}\right)  +\sqrt{2}\left(  \mathbf{Q}_{1}\cdot\mathbf{q}_{1}%
+\mathbf{Q}_{2}\cdot\mathbf{q}_{2}\right)  -\frac{1}{2}\left(  \mathbf{Q}%
_{1}^{2}+\mathbf{Q}_{2}^{2}\right)  \right\}  \label{eq:sorbitalInQ}%
\end{equation}
The choice of parametrization is linked to the basis states one intends to
generate. Associated with our choice of basis (Zernike-Brinkman \cite{kn:ZB}),
we introduce hyperspherical coordinates. The hyperradius and hyperangles, both
for spatial coordinates and for generating parameters, are defined by:
\begin{align}
\rho=\sqrt{\mathbf{q}_{1}^{2}+\mathbf{q}_{2}^{2}},\,  &  q_{1}=\rho\cos
\theta,\,q_{2}=\rho\sin\theta;\nonumber\\
R=\sqrt{\mathbf{Q}_{1}^{2}+\mathbf{Q}_{2}^{2}},\,  &  \,Q_{1}=R\cos
\Theta,\,Q_{2}=R\sin\Theta, \label{eq:HyperCoords}%
\end{align}
Using these, one expands the generating function (\ref{eq:sorbitalInQ}) in HH
functions:
\begin{equation}
\Psi\left(  \mathbf{q_{1},q_{2}}\left\vert \mathbf{Q_{1},Q_{2}}\right.
\right)  =\sum_{\nu}\Psi_{\nu}\left(  \rho,\theta,\mathbf{\check{q}}%
_{1},\mathbf{\check{q}}_{2}\right)  \ \Xi_{\nu}^{\ast}\left(  R,\Theta
,\mathbf{\check{Q}}_{1},\mathbf{\check{Q}}_{2}\right)  \label{eq:GFHS}%
\end{equation}
where the full set of quantum numbers $\nu$ (introduced previously) is
involved in the summation. The oscillator basis functions are%

\begin{equation}
\Psi_{\nu}\left(  \rho,\theta,\mathbf{\check{q}}_{1},\mathbf{\check{q}}%
_{2}\right)  =\mathcal{N}_{n,K\ }\rho^{K}\exp\{-\rho^{2}/2\}\ L_{n}^{K+2}%
(\rho^{2})\ H_{K}^{(l_{1}l_{2})LM}\left(  \theta,\mathbf{\check{q}}%
_{1},\mathbf{\check{q}}_{2}\right)  \label{eq:OscFionHS}%
\end{equation}
and the generator coordinate functions are%

\begin{equation}
\Xi_{\nu}\left(  R,\Theta,\mathbf{\check{Q}}_{1},\mathbf{\check{Q}}%
_{2}\right)  =\mathcal{N}_{n,K}\ R^{K+2n}\ H_{K}^{(l_{1}l_{2})LM}\left(
\Theta,\mathbf{\check{Q}}_{1},\mathbf{\check{Q}}_{2}\right)
\label{eq:GenCoordHS}%
\end{equation}
Here $H$ denotes the HH function%

\begin{align}
H_{K}^{(l_{1}l_{2})LM}\left(  \Theta,\mathbf{\check{Q}}_{1},\mathbf{\check{Q}%
}_{2}\right)   &  =\mathcal{N}_{K}^{(l_{1}l_{2})LM}\Phi_{K}^{(l_{1}l_{2}%
)}\left(  \Theta\right)  \left\{  \operatorname*{Y}\nolimits_{l_{1}%
}(\mathbf{\check{Q}}_{1})\times\operatorname*{Y}\nolimits_{l_{2}%
}(\mathbf{\check{Q}}_{2})\right\}  _{LM}\nonumber\\
\Phi_{K}^{(l_{1}l_{2})}\left(  \Theta\right)   &  =\left(  \cos\Theta\right)
^{l_{1}}\ \left(  \sin\Theta\right)  ^{l_{2}}\ P_{\frac{K-l_{1}-l_{2}}{2}%
}^{l_{2}+\frac{1}{2},l_{1}+\frac{1}{2}}\,(\cos2\Theta) \label{eq:HSFion}%
\end{align}
From (\ref{eq:GenCoordHS}), one easily deduces the procedure for selecting
basis functions with fixed quantum numbers $\nu=\{n,K,\left(  l_{1}%
l_{2}\right)  LM\}$. One has to differentiate the generating function $\left(
K+2n\right)  $-times with respect to $R$ and then to set $R=0$. After that one
has to integrate over $\Theta$ with the weight $\Phi_{K}^{(l_{1}l_{2})}$ to
project onto the hypermomentum $K$; one has to integrate over unit vectors
$\mathbf{\check{Q}}_{1}$ and $\mathbf{\check{Q}}_{2}$ with weights
$\operatorname{Y}_{l_{1}m_{1}}(\mathbf{\check{Q}}_{1})$ and $\operatorname{Y}%
_{l_{2}m_{2}}(\mathbf{\check{Q}}_{2})$ to project onto partial angular momenta.

The calculation of matrix elements with the generating function method is thus
a two-step process. The first step is the calculation of the generating
function for the operator involved. Usually this is accomplished with
analytical techniques. The second step is the expansion of the generating
function w.r.t. the generator coordinates. Either explicit differentiation or
recurrence relations can be used to derive expressions for individual matrix
elements \cite{kn:AM_AJP}. In any case, the work involved is straightforward
but extremely tedious; both approaches are best implemented using algebraic
manipulation software such as Mathematica or Maple.

A further explicit presentation of the calculation of matrix elements for
overlap and Hamiltonian is beyond the scope of the current paper because of
the highly technical details involved. We refer to \cite{kn:ITP+RUCA1} for
explicit details for the three-cluster case.

\subsection{The Modified JM approach}

The numerical solution of the JM equations critically depends on an
appropriate choice of $N$, separating the internal from the external region.
The determining factor in this is the behavior of the potential energy matrix
elements which can be slowly decreasing as a function of $n$. If this is the
case, a sufficiently large value of $N$ has to be chosen.

In the case of three-cluster systems it is known from literature
\cite{kn:Fedorov94} that the potential asymptotically behaves as $1/\rho^{3}$
in the hyperradius, with a corresponding effect on the matrix elements. It
will be shown later that the asymptotic form of the effective potential in our
calculation indeed follows this behavior. Potentials with an asymptotic tail
$1/\rho^{3}$ dramatically affect the phase shift in the low energy region
\cite{kn:Calogero,kn:Babikov} and special care should be taken to get
convergent results. This usually requires more than merely choosing a
sufficiently large value of $N$.

Another factor that aversely affects convergence is the choice of a single
basis (i.e. a fixed oscillator length $b$) for both the cluster and relative
motion basis functions. This choice has been made to make the calculation of
matrix elements manageable, and within acceptable execution time. The value
chosen for $b$ is usually fixed so that the frozen clusters have appropriate
physical properties (e.g. ground-state energy). It is indeed important that
the individual fragments are well described, as their properties cannot be
influenced by the solution of the dynamical equations which only involves
their relative behavior. This choice of basis is often far from optimal for a
convergent calculation.

To account for these convergence problems we have considered the MJM approach,
developed in \cite{2002PhRvL..88a0404V,kn:VA_PR}. In this approach we consider
a semi-classical approximation for the matrix elements of the potential that
is incorporated in the three-term recursion relation. We so introduce an
intermediate region between the pure interaction and asymptotic regions. In
this intermediate region a modified asymptotic solution holds, obtained from
the modified recursion relation, and incorporating potential effects. In
particular a long asymptotic tail of the potential can be taken into account
in a satisfactory way. The matching point for the boundary condition,
originally positioned at the border between interaction and asymptotic
regions, is brought down to the border between the near interaction and
intermediate regions. This dramatically reduces the dimensions of the
remaining set of matrix equations, as well as improves overall convergence by
taking into account the asymptotic tail of the potential energy.

\section{Resonances in the MJM $^{6}He$ and $^{6}Be$ in the three-cluster
approach}

\label{sect:ThreeClResonances}

In this section we consider the continuum obtained in a three-cluster MJM
calculation for the 6-particle nuclei $^{6}He$ and $^{6}Be$, determined by the
three-cluster configurations $\alpha+n+n$ and $\alpha+p+p$. Our objective is
to highlight the characteristics of MJM three-cluster calculations, and to
produce accurate results for the astrophysically relevant resonances in
$^{6}He$ and $^{6}Be$.

As the MJM for three-cluster scattering leads to a set of coupled-channel
Schr\"{o}dinger equations (see section ...), we will transform the $S$-matrix
to diagonal form. This is usually referred to as the eigenchannel
representation of the $S$-matrix. We will derive the position and width of the
resonances eigenphase shifts $\delta$ as a function of energy. To be precise,
we use the conditions
\begin{equation}
\frac{d^{2}\delta}{dE^{2}}|_{E=E_{r}}=0,\quad\Gamma=2\left[  \frac{d\delta
}{dE}\right]  ^{-1}|_{E=E_{r}} \label{eq:ResConds}%
\end{equation}

We particularly focus on the properties of the low-lying 0$^{+}$-and 2$^{+}%
$-resonances in $^{6}Be$ and the 2$^{+}$-resonance state in $^{6}He$.

Several models and methods have already been applied to investigate the
resonance states of $^{6}He$ and $^{6}Be$. In the series of papers
\cite{kn:Zhuk93,kn:Dani91,kn:Dani97}, the nuclei were considered as
three-particle systems, neglecting antisymmetrization. These authors used an
effective interaction between the alpha-particle and a nucleon, and a
nucleon-nucleon ($NN$) realistic potential between the two nucleons. The HH's
Method was used to describe the bound as well as the three-particle continuum
states. In \cite{kn:Csoto94} the Complex Scaling Method was used to calculate
the characteristic parameters of the resonance states in $^{6}He $ and
$^{6}Be$. This was done within a three-cluster model, in which full
antisymmetrization was taken into account, and the interactions between
clusters was obtained from the Minnesota $NN$-potential. In
\cite{kn:CSM-tanaka} it was demonstrated that the method of continuation in
the coupling constant (CCCM), used for the $\alpha+N+N$ configurations in
$^{6}He$ and $^{6}Li$, reproduces results, which are very close to those
obtained by the CSM. We will compare MJM results to each of these methods. The
MJM formulation leads to a good convergence of the computations, even for the
(important) Coulomb contribution. An important renormalization effect on the
irregular asymptotic solution still remains a restrictive factor in the
reduction of the number of basis states in some cases.

\subsection{Details of the calculation}

The Volkov potential \cite{kn:Volk65} has been used to describe $NN$
interaction in all calculations of this section. According to
\cite{kn:Vasil96,kn:Vasil97} this provides an acceptable description for
$^{6}He$ within the three-cluster model. The effective Volkov potential
consists of central forces only, so that total angular momentum $L$, total
spin $S$ and total isospin $T$ are good quantum numbers. We will therefore
consider only three-cluster configurations $\alpha+N+N$ with $S=0$ and $T=1$.
This is known to be the most prominent spin-isospin state for the resonances
of interest in the nuclei considered in this work. The Coulomb interaction has
also been included, as it is to a great extent responsible for reproducing the
0$^{+}$ resonance state in $^{6}Be$.

The oscillator radius $b$, associated with the basis functions for the nuclear
state, is the only free parameter for the MJM calculation. It's value was
chosen to optimize the ground-state energy of the $\alpha$-particle, and
equals $b=1.37$ fm.

As explained before, we have a choice of Jacobi coordinate systems and this
can be exploited to simplify the calculations. For a cluster configuration
$\alpha+N+N$ one can consider the two Jacobi configurations displayed in
figure \ref{fig:figure1}. The first (referred to as the \textquotedblleft%
4+2\textquotedblright-configuration) is the\ most appropriate for the current
calculation. Selection rules significantly reduce the number of basis
functions. Indeed, quantum numbers $S=0,T=1$ coincide for the full six-nucleon
system and the two-nucleon subsystem ($N+N$). Hence the relative motion wave
function must be an even function in the coordinate $q_{1}$ of this Jacobi
system. This means that only even angular momenta $l_{1}$ have to be
considered. Moreover, for positive parity states only even values of $l_{2}$,
and for negative parity states only odd values of $l_{2}$ should be taken into
account. With the second Jacobi configuration from figure \ref{fig:figure1}%
\ (the \textquotedblleft5+1\textquotedblright\ one), these constraints are
hard to meet and the full oscillator basis would have to be used.%
\begin{table}[tbp] \centering
\caption{Number ($N_{h}$) and accumulated number ($N_{c}$) of Hyperspherical Harmonics
up to $K=10$}
\begin{tabular}
[c]{|c|c|c|c|c|c|c|c|}\hline
$L^{\pi}=0^{+}$ & $K$ & 0 & 2 & 4 & 6 & 8 & 10\\\hline
& $N_{h}$ & 1 & 1 & 2 & 2 & 3 & 3\\
& $N_{c}$ & 1 & 2 & 4 & 6 & 9 & 12\\\hline
$L^{\pi}=2^{+}$ & $K$ & 0 & 2 & 4 & 6 & 8 & 10\\\hline
& $N_{h}$ & - & 2 & 3 & 5 & 6 & 8\\
& $N_{c}$ & - & 2 & 5 & 10 & 16 & 24\\\hline
\end{tabular}
\label{tab:HH state}
\end{table}%

Table \ref{tab:HH state} shows the number of HH's ($N_{h}$) of given
hypermomentum $K$ for $L^{\pi}=0^{+}$ and $L^{\pi}=2^{+}$ in the ``4+2''
Jacobi configuration. This table also indicates the total number of HH's so
far ($N_{c}$) with $K=K_{\min},$ $K_{\min}+2,\ldots,K_{\max}$, i.e. it shows
the number of channels for a given $K_{\max}=K$.

\subsubsection{Overlap matrix elements}

In the interaction region we apply full antisymmetrization, i.e. also
inter-cluster anti-symmetrization. Here we look at its effect on positioning
the boundary between the internal (interaction) region and the asymptotic
region. No particle exchanges should occur in the asymptotic region. In the
MJM the antisymmetrization effects can make themselves felt through the
overlap and potential matrix elements. We use the notations of
\cite{kn:ITP+RUCA1} for the overlap matrix elements, and in particular the
shorthand notation $\nu_{0}$\ is used for the set $(l_{1}l_{2})LM$\ of quantum
numbers:
\begin{equation}
\left\langle n,\left(  K,\nu_{0}\right)  \left\vert \widehat{\mathcal{A}%
}\right\vert n^{\prime},\left(  K^{\prime},\nu_{0}\right)  \right\rangle
\label{eq:OvlapMatrixElem}%
\end{equation}
Non-zero matrix elements (\ref{eq:OvlapMatrixElem}) can be obtained from
states within the same many-particle oscillator shell only. As the oscillator
shells in the Hyperspherical description are characterized by $N=2n+K$ , the
selection rule becomes $2n+K=2n^{\prime}+K^{\prime}$.%
\begin{figure}
[ptb]
\begin{center}
\includegraphics[
trim=0.000000in 0.102347in 0.000000in 0.510910in,
natheight=8.253800in,
natwidth=11.681000in,
height=7.5718cm,
width=10.803cm
]%
{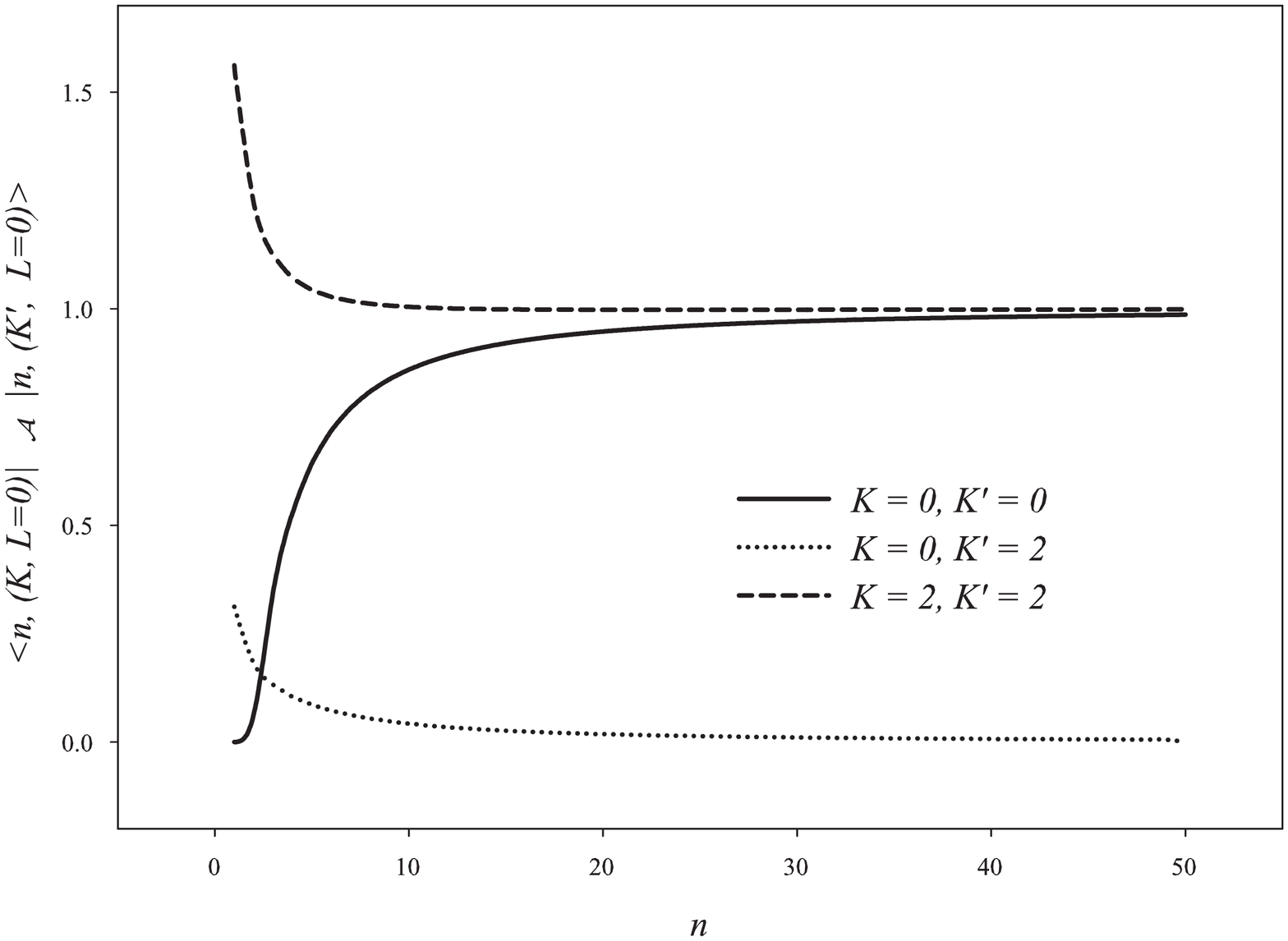}%
\caption{Matrix elements of the antisymmetrization operator for the $0^{+}%
$-state in $^{6}He$ and $^{6}Be$}%
\label{fig:figure2}%
\end{center}
\end{figure}

In figure \ref{fig:figure2}\ overlap matrix elements diagonal in $n$ for $L=0
$ and hypermomenta $K=0$ and $K=2$ are shown for the six-nucleon three-cluster
system. One notices from this figure that the Pauli principle involves
oscillator states of at least the 25 lowest shells, for both diagonal and
off-diagonal matrix elements in $K$. The antisymmetrization effects are
visible in the deviation from unity for the diagonal matrix elements, and the
deviation from zero of the off-diagonal (in $K$) elements. These effects
decrease monotonically with higher $n$.%
\begin{figure}
[ptb]
\begin{center}
\includegraphics[
trim=0.000000in 0.102347in 0.000000in 0.510910in,
natheight=8.253800in,
natwidth=11.681000in,
height=7.5718cm,
width=10.8008cm
]%
{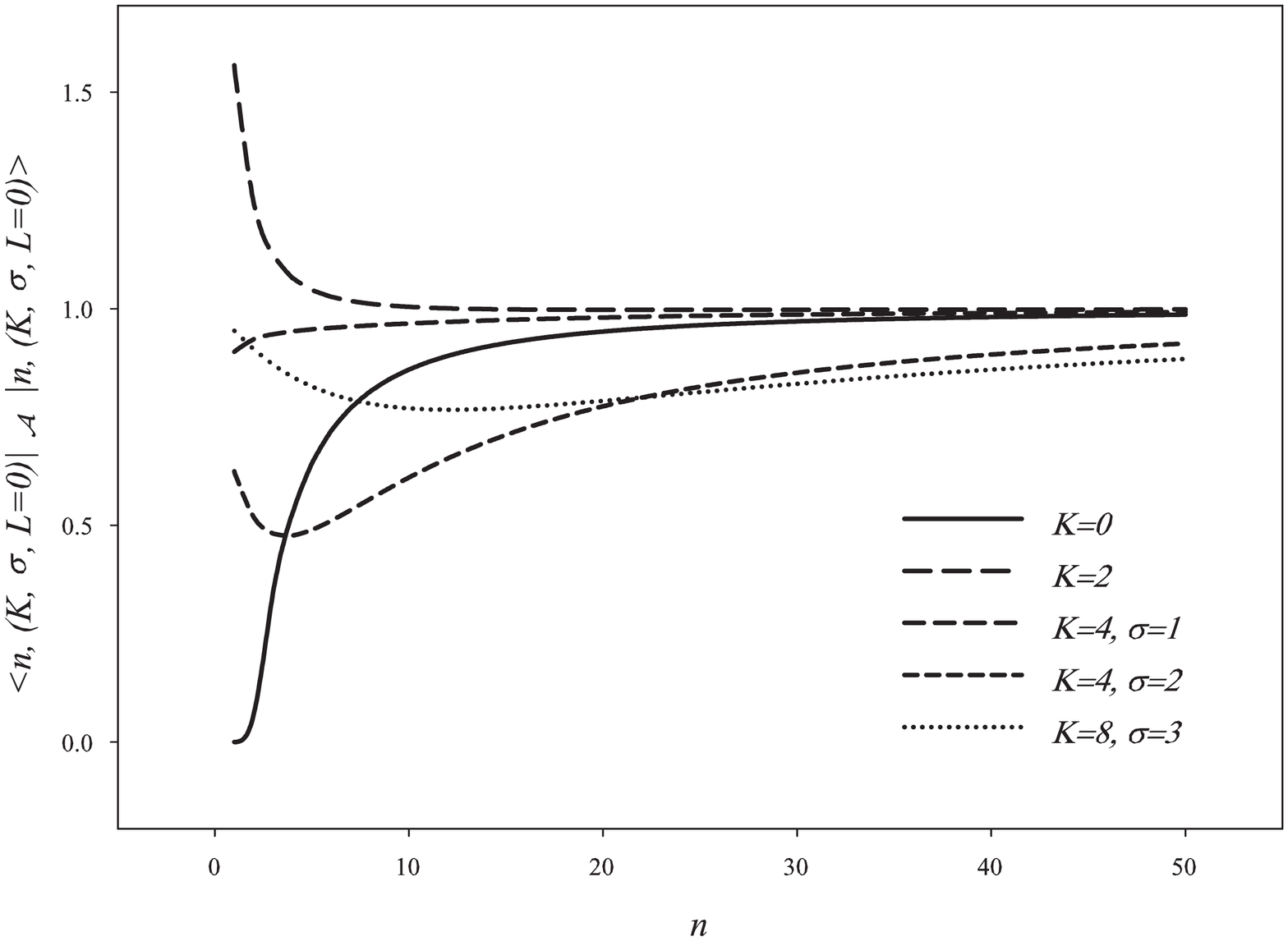}%
\caption{Matrix elements of the antisymmetrization operator for the $0^{+}%
$-state in $^{6}He$ and $^{6}Be$}%
\label{fig:figure3}%
\end{center}
\end{figure}

In figure \ref{fig:figure3} we compare matrix elements for $L=0$ diagonal in
$n$ and $K$ for some of the $K$-values, with $\sigma$ a multiplicity quantum
number for states with identical $K$. Only those matrix elements where the
Pauli principle effect is most prominent have been shown. Some states with
$K=4$ and $K=8$ are affected more strongly by antisymmetrization than others.
To understand this we note that the Hyperspherical angles (corresponding to
the hyperangular quantum numbers $K,l_{1},l_{2},LM$) define the most probable
triangular shape and orientation in space of the three-cluster system. The
HH's with $K=4,\sigma=2$ (characterized by $l_{1}=l_{2}=2$) and $K=8,\sigma=3$
(characterized by $l_{1}=l_{2}=4$) seem to describe a triangular shape where
one of the nucleons is very close to the $\alpha$-particle.

For larger $K$-values the probability to find all clusters close to one
another within a hypersphere of fixed radius $\rho$ decreases, and one can
expect that HH's with large values of $K$ will play a diminishing role in the
calculations.%
\begin{figure}
[ptb]
\begin{center}
\includegraphics[
trim=0.000000in 0.102347in 0.000000in 0.510910in,
natheight=8.253800in,
natwidth=11.681000in,
height=7.5718cm,
width=10.8008cm
]%
{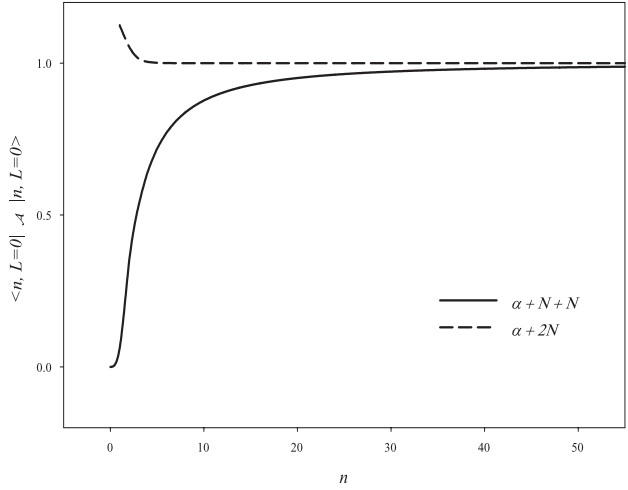}%
\caption{Matrix elements of the antisymmetrization operator for the 0$^{+}%
$-state of $^{6}He$ and $^{6}Be$ in the three-cluster $\alpha+N+N$ ($K=0$) and
the two-cluster configuration $\alpha+2N$ }%
\label{fig:figure4}%
\end{center}
\end{figure}

It interesting to compare overlap matrix elements for the three-cluster
configuration $\alpha+N+N$ with those of the two-cluster configurations
$\alpha+2N$ (such as $\alpha+d$ in $^{6}Li$ or $\alpha+2n$ in $^{6}He$). This
comparison is shown in figure \ref{fig:figure4}, and it indicates that the
Pauli principle has a much larger \textquotedblleft range\textquotedblright%
\ in the three-cluster than in the two-cluster configuration.

\subsubsection{Potential matrix elements}

To investigate the Pauli effect on the potential energy of the three-cluster
configuration $\alpha+N+N$ we compare the potential matrix elements with
hypermomentum $K=0$ with full antisymmetrization against those in the folding
approximation, where antisymmetrization between clusters is neglected.
\begin{figure}
[ptb]
\begin{center}
\includegraphics[
trim=0.000000in 0.102347in 0.000000in 0.510910in,
natheight=8.253800in,
natwidth=11.681000in,
height=7.5718cm,
width=10.7986cm
]%
{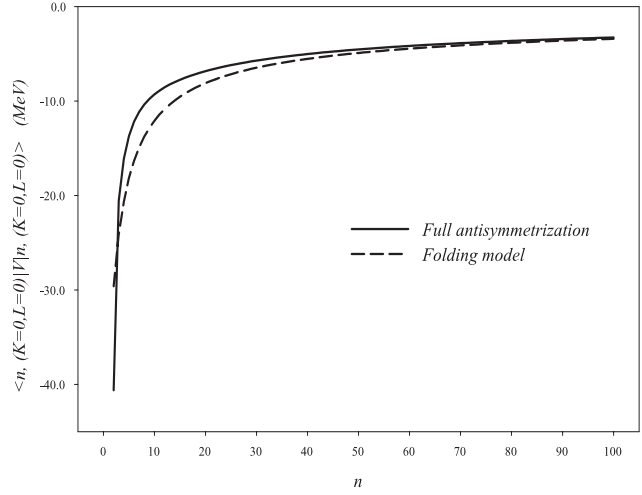}%
\caption{Diagonal potential matrix elements for the $0^{+}$-state of $^{6}Be$
and $^{6}He$ ($K=0$) with full antisymmetrization, and in the folding model}%
\label{fig:figure5}%
\end{center}
\end{figure}
Figure \ref{fig:figure5} shows the diagonal potential matrix elements diagonal
in $n$ for $K=0$ and figure \ref{fig:figure6} shows the matrix elements along
a fixed row $n=50$ for $K=0$. One notes that the folding model results are
very close to the fully antisymmetrized ones, especially for larger $n$.%
\begin{figure}
[ptbptb]
\begin{center}
\includegraphics[
trim=0.000000in 0.102347in 0.000000in 0.510910in,
natheight=8.253800in,
natwidth=11.681000in,
height=7.5718cm,
width=10.6097cm
]%
{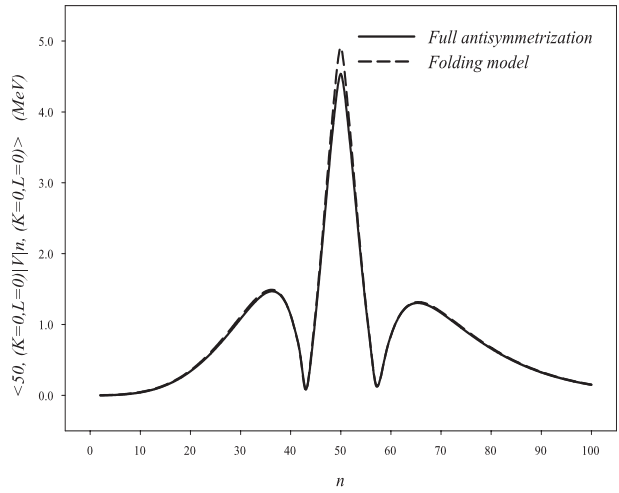}%
\caption{Off-diagonal potential matrix elements for the 0$^{+}$-state of
$^{6}Be$ and $^{6}He$ ($K=0$) with full antisymmetrization, and in the folding
model}%
\label{fig:figure6}%
\end{center}
\end{figure}
\begin{figure}
[ptbptbptb]
\begin{center}
\includegraphics[
trim=0.000000in 0.102347in 0.000000in 0.510910in,
natheight=8.253800in,
natwidth=11.681000in,
height=7.5718cm,
width=10.803cm
]%
{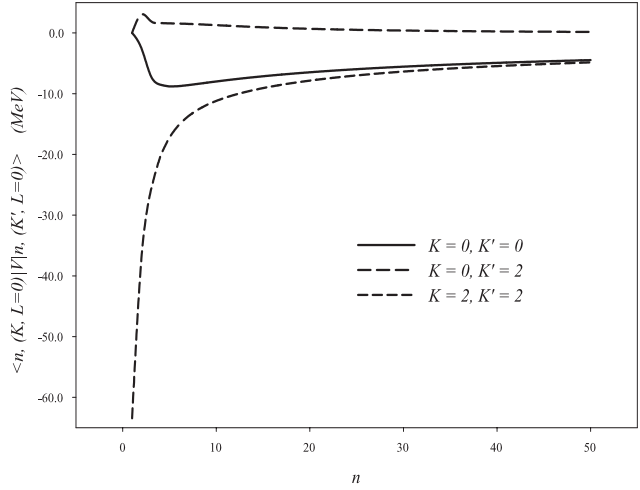}%
\caption{Diagonal potential matrix elements for the 0$^{+}$-state of $^{6}Be $
and $^{6}He$, with full antisymmetrization}%
\label{fig:figure7}%
\end{center}
\end{figure}

In figure \ref{fig:figure7}\ we also display the potential matrix elements
between states of the two lowest values of hypermomentum $K=0$ and $K=2$. One
notices that the $K=2$ contribution is the largest. The potential energy for
$K=0$ is relatively small, and so is the coupling between $K=0$ and $K=2$ states.

The main conclusion is that, in the asymptotic region, the \textquotedblleft
exact\textquotedblright\ potential energy can be substituted with the folding
approximation. It leads to a considerable reduction in computational effort.
We are led to the following setup for three--cluster calculations. In the
internal region, consisting of states of the lower oscillator shells and with
a large probability to find the clusters close to one another, the fully
antisymmetrized potential energy is used. In the asymptotic region, where the
average distance between clusters is large, we use the folding model
potential.%
\begin{figure}
[ptb]
\begin{center}
\includegraphics[
trim=0.000000in 0.102347in 0.000000in 0.510910in,
natheight=8.253800in,
natwidth=11.681000in,
height=7.5718cm,
width=10.803cm
]%
{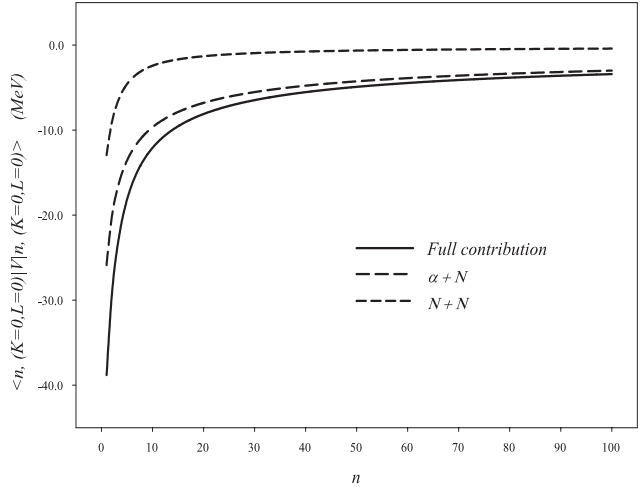}%
\caption{Folding model contributions in the diagonal potential matrix elements
of the energy operator for the $0^{+}$-state of $^{6}Be$ and $^{6}He$ ($K=0$)
}%
\label{fig:figure8}%
\end{center}
\end{figure}

The folding model provides additional insight in the structure of the
interaction matrix. It is well known that neither the $\alpha+n$ nor the $n+n$
interaction can create a bound state in the corresponding subsystems of
$^{6}He$. Only a full three--cluster configuration $\alpha+n+n$ contains the
necessary conditions to create a bound state. In the folding model the total
potential energy is a sum of contributions from the three interacting pairs:
the $\alpha$--particle and the first neutron, the $\alpha$--particle and the
second neutron and the neutron-neutron pair. For $K=0$ the first two
contributions are identical, and we only have to consider the $\alpha+n$ and
the $n+n$ pairs. Figure \ref{fig:figure8} shows the diagonal matrix element
contribution of both components, and one notices that $\alpha+n$ represents
the main contribution.

\subsubsection{The effective charge.}

When Coulomb forces are taken into account, one needs to determine the
effective charge in order to properly solve the MJM equations. The effective
charge unambiguously defines the effective Coulomb interaction in each channel
as well as the coupling between different channels. Using the approach
suggested in section \ref{sect:AsympSolCoordRep}, the effective charges for
the $0^{+}$- and $2^{+}$-states of $^{6}Be$ were calculated. Part of the
corresponding matrices of $\left\Vert Z_{K}^{K^{\prime}}\right\Vert $ are
displayed in tables \ref{tab:table1} and \ref{tab:table2} respectively. One
notices that the diagonal matrix elements are much larger than the
off-diagonal ones. This justifies the approximation \cite{kn:ITP+RUCA1}\ of
disregarding the coupling of the channels in the asymptotic region.%
\begin{table}[tbp] \centering
\caption{Effective charge matrix for the 0$^{+}$-state in $^{6}Be$}
\begin{tabular}
[c]{|c|c|c|c|c|}\hline
$K;l_{1},l_{2}$ & $0;0,0$ & $2;0,0$ & $4;0,0$ & $4;2,2$\\\hline
$0;0,0$ & 7.274 & 0.006 & -0.129 & 1.414\\\hline
$2;0,0$ & 0.006 & 7.146 & -0.436 & 0.314\\\hline
$4;0,0$ & -0.129 & -0.436 & 7.428 & -0.877\\\hline
$4;2,2$ & 1.414 & 0.314 & -0.877 & 9.098\\\hline
\end{tabular}
\label{tab:table1}
\end{table}
\begin{table}[tbp] \centering
\caption{Effective charge matrix for the  2$^{+}$-state in $^{6}Be$}
\begin{tabular}
[c]{|l|l|l|l|l|l|}\hline
$K;l_{1},l_{2}$ & $2;2,0$ & $2;0,2$ & $4;2,0$ & $4;0,2$ & $4;2,2$\\\hline
$2;2,0$ & 7.253 & 0.400 & -0.224 & 0.546 & -0.751\\\hline
$2;0,2$ & 0.400 & 7.244 & -0.309 & -0.004 & -0.601\\\hline
$4;2,0$ & -0.224 & -0.309 & 6.942 & -0.186 & 0.431\\\hline
$4;0,2$ & 0.546 & -0.004 & -0.186 & 7.694 & -0.671\\\hline
$4;2,2$ & -0.751 & -0.601 & 0.431 & -0.671 & 7.345\\\hline
\end{tabular}
\label{tab:table2}
\end{table}%

It is interesting to compare the three-cluster effective charge\ with the
effective charge in the two-cluster configuration. For the latter we can
write
\[
Z=Z_{1}Z_{2}e^{2}\sqrt{\frac{A_{1}A_{2}}{A_{1}+A_{2}}}%
\]
where $A_{1}$ and $Z_{1}$\ ($A_{2}$ and $Z_{2}$) are the respective mass and
charge of both clusters. For the configuration $\alpha+2p$ in $^{6}Be$, we
then obtain an effective charge
\[
Z=\frac{8}{\sqrt{3}}e^{2}\simeq6.65
\]
which is independent on the angular momentum of the system. One notice that
the two-cluster effective charge is close to the diagonal matrix elements of
$\left\Vert Z_{K}^{K^{\prime}}\right\Vert $. We can assume that, if\ in one of
the three-cluster channels the effective charge is very close to the
two-cluster one, it could indicate that the two protons move as an aggregate
in the asymptotic region. For the $2^{+}$-state we observe at least one
channel with this property, carrying the labels $K=4$, $l_{1}=2,$ $l_{2}=0$.

\subsection{Results}

\subsubsection{Definition of the model space}

The model space for the current calculations is primarily determined by the
total number of HH's in the internal and external region, and the number of
oscillator states. Different sets of HH's can be used in the internal and
asymptotic regions. An extensive set of HH's in the internal region will
provide a well correlated description of the three-cluster system due to the
coupling between states with different hypermomentum. The HH's in the
asymptotic region, which are exactly (without Coulomb) or nearly exactly
(Coulomb included) decoupled, are responsible for the richness in decay possibilities.

In the current paper we restricted both internal and asymptotic HH sets
maximal hypermomentum value $K_{\max}^{(i)}=K_{\max}^{(a)}=10$. By extending
the respective subspaces up to these maximal values, we obtain a fair
indication of convergence.%
\begin{figure}
[ptb]
\begin{center}
\includegraphics[
trim=0.000000in 0.102347in 0.000000in 0.510910in,
natheight=8.253800in,
natwidth=11.681000in,
height=7.5696cm,
width=10.8623cm
]%
{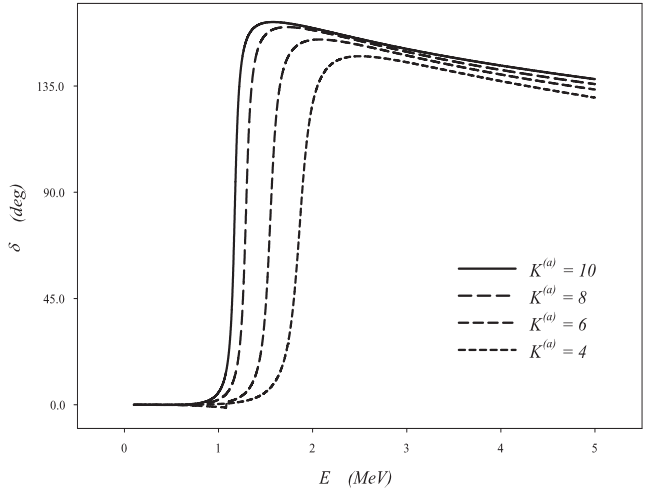}%
\caption{Eigenphase shifts for the $0^{+}$-state of $^{6}Be$ for different
$K_{max}^{(a)}$}%
\label{fig:figure9}%
\end{center}
\end{figure}

We fixed the matching point between internal and asymptotic region at $N=50$.
It is based on the observations of the previous sections, i.e. (1) it is
sufficiently large to allow the Pauli principle its full impact, and (2) it is
large enough for the semi-classical approximations in the MJM to be valid.

The resonance parameters are obtained from the eigenphase shifts obtained from
the eigenchannel representation (diagonalized form) of the $S$-matrix.

\subsubsection{Convergence study}

We first consider the influence of the number of asymptotic channels. We do
this by calculating the position and width of the 0$^{+}$-state in $^{6}Be$
using successively larger asymptotic subspaces. The first calculation includes
all HH's up to $K_{\max}^{(i)}=8$ and extends the asymptotic subspace from
$K_{\max}^{(a)}=0$ to $K_{\max}^{(a)}=8$. The second calculation includes all
HH's up to $K_{\max}^{(i)}=10$ and extends the asymptotic subspace from
$K_{\max}^{(a)}=0$ to $K_{\max}^{(a)}=10$.
\begin{table}[tbp] \centering
\caption{0$^{+}$-resonance parameters for $^6$Be, varying
$K_{max}^{(a)}$ ($K_{max}^{(i)}=8$)}%
\begin{tabular}
[c]{|l|l|l|l|l|l|}\hline
$K_{\max}^{(a)}$ & 0 & 2 & 4 & 6 & 8\\\hline
$E$, MeV & 1.434 & 1.314 & 1.304 & 1.298 & 1.292\\
$\Gamma$, MeV & 0.075 & 0.082 & 0.084 & 0.085 & 0.087\\\hline
\end{tabular}
\label{tab:Conv_vs_asy_8}%
\end{table}
\begin{table}[tbp] \centering
\caption{0$^{+}$-resonance parameters for $^6$Be, varying
$K_{max}^{(a)}$ ($K_{max}^{(i)}=10$)}%
\begin{tabular}
[c]{|c|c|c|c|c|c|c|}\hline
$K_{\max}^{(a)}$ & 0 & 2 & 4 & 6 & 8 & 10\\\hline
$E$, MeV & 1.324 & 1.204 & 1.192 & 1.184 & 1.176 & 1.172\\
$\Gamma$, MeV & 0.068 & 0.069 & 0.071 & 0.071 & 0.073 & 0.072\\\hline
\end{tabular}
\label{tab:Conv_vs_asy_10}%
\end{table}
The corresponding results are shown in tables \ref{tab:Conv_vs_asy_8} and
\ref{tab:Conv_vs_asy_10}. One learns from these results that a sufficient rate
of convergence has been obtained. Figure \ref{fig:figure9} displays the first
eigenphase shift as a function of energy for $K_{\max}^{(i)}=10$ and a choice
of $K_{\max}^{(a)}$ values from which results in the tables are derived.%
\begin{table}[tbp] \centering
\caption{0$^{+}$-resonance parameters for $^6$Be, varying $K_{max}%
^{(i)}$ ($K_{max}^{(a)}=0$)}%
\begin{tabular}
[c]{|c|c|c|c|c|c|c|}\hline
$K_{\max}^{(i)}$ & 0 & 2 & 4 & 6 & 8 & 10\\\hline
$E$, MeV & - & 2.408 & 2.020 & 1.688 & 1.434 & 1.324\\
$\Gamma$, MeV & - & 0.147 & 0.129 & 0.097 & 0.075 & 0.068\\\hline
\end{tabular}
\label{tab:Conv_vs_int}%
\end{table}%

Whereas the inclusion of higher hypermomenta in the asymptotics shows a fast
and monotonic convergence, it is also clear from these results that a
sufficient number of HH's has to be used for a correct description of the
correlations in the internal state. To support this conclusion, we performed a
calculation, again for the 0$^{+}$-state in $^{6}Be$, in which only one HH
with value $K^{(a)}=0$ was used. In the internal region the number of HH's was
varied from $K^{(i)}=0$ up to $K^{(i)}=10$. These results appear in table
\ref{tab:Conv_vs_int} and corroborate the previous conclusion. In particular
they indicate that the effective potential obtained with the $K^{(i)}=0$ HH
only, is unable to produce a resonance. Only after including a $K=2$ HH does
the resonance appear. Further inclusion of higher HH states then lead to an
acceptable convergence in a monotonically decreasing fashion for both position
and width of the resonance.%
\begin{table}[tbp] \centering
\caption{Second resonance state parameters, obtained by MJM, HHM and
CSM}%
\begin{tabular}
[c]{|l|l|cc|cc|cc|}\hline
Nucleus & $L^{\pi}$ & \multicolumn{2}{c|}{MJM} &
\multicolumn{2}{l|}{HH\cite{kn:Dani91}} &
\multicolumn{2}{l|}{CSM\cite{kn:CSM-Aoyama1} \cite{kn:CSM-Aoyama2}}\\\hline
&  & $E$, MeV & $\ \Gamma$, MeV & $E$, MeV & $\Gamma$, MeV & $E$, MeV &
$\Gamma$, MeV\\\hline
$^{6}He$ & 0$_{2}^{+}$ & 2.1 & 4.3 & 5.0 & 6.0 & 3.9 & 9.4\\\hline
$^{6}He$ & 2$_{2}^{+}$ & 3.7 & 5.0 & 3.3 & 1.2 & 2.5 & 4.7\\\hline
$^{6}Be$ & 0$_{2}^{+}$ & 3.5 & 6.1 &  &  &  & \\\hline
$^{6}Be$ & 2$_{2}^{+}$ & 5.2 & 5.6 &  &  &  & \\\hline
\end{tabular}
\label{tab:second_res}%
\end{table}
\begin{table}[tbp] \centering
\caption{Resonance state parameters for $^6He$ and $^6Be$, obtained
by MJM, HHM, CSM and CCCM}%
\begin{tabular}
[c]{|l|cc|cc|cc|}\hline
\  & \multicolumn{2}{c|}{$^{6}He;L^{\pi}=2^{+}$} & \multicolumn{2}{c|}{$^{6}%
Be;L^{\pi}=0^{+}$} & \multicolumn{2}{c|}{$^{6}Be;L^{\pi}=2^{+}$}\\\hline
Method & $E$, MeV & $\Gamma$, MeV & $E$, MeV & $\Gamma$, MeV & $E$, MeV &
$\Gamma$, MeV\\\hline
MJM & 1.490 & 0.168 & 1.172 & 0.072 & 3.100 & 0.798\\\hline
HHM\cite{kn:Dani91} & 0.75 & 0.04 &  &  &  & \\\hline
CSM \cite{kn:Csoto94} & 0.74 & 0.06 & 1.52 & 0.16 & 2.81 & 0.87\\\hline
CCCM\cite{kn:CSM-tanaka} & 0.73 & 0.07 &  &  &  & \\\hline
\end{tabular}
\label{tab:Theory}%
\end{table}%

\subsubsection{Comparisons}

\begin{figure}
[ptb]
\begin{center}
\includegraphics[
trim=0.000000in 0.103173in 0.000000in 0.513386in,
natheight=8.253800in,
natwidth=11.681000in,
height=7.5256cm,
width=10.8184cm
]%
{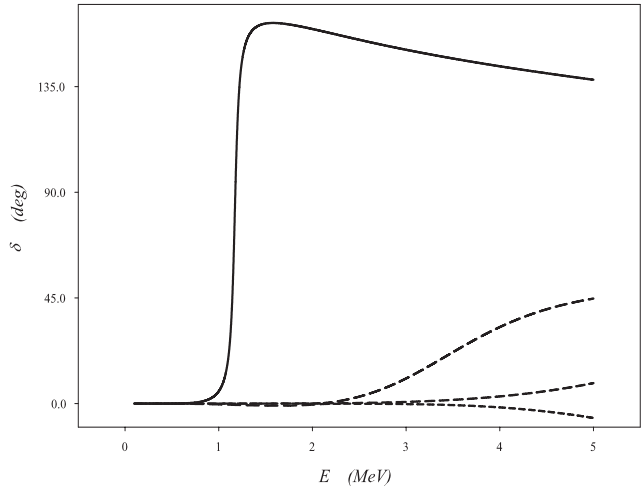}%
\caption{The\ eigenphases for the $0^{+} $-state in $^{6}Be$}%
\label{fig:figure10}%
\end{center}
\end{figure}

In figure \ref{fig:figure10} we display eigenphase shifts for $L^{\pi}=0^{+} $
in $^{6}Be$ in the full calculations, i.e. with the maximal number of internal
and asymptotic HH's. One notices that the first $0^{+}$-resonance state of
$^{6}Be$ appears in the first eigenchannel, and that a second (broad)
resonance at a higher energy is created in the second eigenchannel (see table
\ref{tab:second_res} for details on this resonance). The phase shifts in the
higher eigenchannels show a smooth behavior as a function of energy without a
trace of resonances in the energy range that we consider.%
\begin{table}[tbp] \centering
\caption{Comparison of resonance state parameter in $^{6}He$ and $^{6}Be$
between MJM and experiment.}%
\begin{tabular}
[c]{|c|cc|cc|}\hline
\  & \multicolumn{2}{c|}{MJM} & \multicolumn{2}{c|}{Experiment
\cite{kn:Ajze88}}\\\hline
& $E$, MeV & $\Gamma$, MeV & $E$, MeV & $\Gamma$, MeV\\\hline
$^{6}He$; $L^{\pi}=2^{+}$ & 1.490 & 0.168 & 0.822$\pm$0.025 & 0.133$\pm
$0.020\\\hline
$^{6}Be$; $L^{\pi}=0^{+}$ & 1.172 & 0.072 & 1.371 & 0.092$\pm$0.006\\\hline
$^{6}Be$; $L^{\pi}=2^{+}$ & 3.100 & 0.798 & 3.04$\pm$0.05 & 1.16$\pm
$0.06\\\hline
\end{tabular}
\label{tab:AM+exprm}%
\end{table}%

Table \ref{tab:Theory} compares the results of this work to those obtained in
other calculations, in particular by CSM \cite{kn:Csoto94}, by HHM
\cite{kn:Dani91} and by CCCM \cite{kn:CSM-tanaka}. In table \ref{tab:AM+exprm}
our results are compared with the experimental data available from
\cite{kn:Ajze88}. The agreement with the experimental energy and width of the
resonant states is reasonable. The difference between the experimental and
calculated energies of the 2$^{+}$-resonance states in both $^{6}He$ and
$^{6}Be$ are probably due to the lack on LS-forces in the present calculations.

It has been pointed out in \cite{kn:Dani91} and \cite{kn:Csoto94}, that the
barrier created by the three-cluster configuration is sufficiently high and
wide to accommodate two resonance states, the second one usually being very
broad. At first sight, this could be ascribed to an artifact of the HH Model
since large values of hypermomentum $K$ create a substantial centrifugal
barrier. However the CSM calculations \cite{kn:CSM-Aoyama1,kn:CSM-Aoyama2},
which do not use HH's, also reveal such resonances. Our calculations now also
confirm the existence of a second very broad resonance. A comparison of the
resonance parameters with those of the HH-method \cite{kn:Dani91} and of the
CSM \cite{kn:Csoto94} is shown in table \ref{tab:second_res}. The differences
are most probably due to the different descriptions of the system, as well as
to the difference in $NN$-forces.

\section{The $^{3}H(^{3}H,2n)^{4}He$ and $^{3}He(^{3}He,2p)^{4}He$ reactions
in the MJM cluster approach}

A second application that shows the strength of the MJM approach is connected
to the nuclear reaction $^{3}He(^{3}He,2p)^{4}He$ which contributes for 89\%
to the $pp$-chain of nuclear synthesis, and thus is of particular interest to
astrophysics. We will couple the two- and three-cluster descriptions to
calculate the reaction properties, by considering a two-cluster entrance, and
a three-cluster exit channel

The theoretical analysis of the $^{3}He(^{3}He,2p)^{4}He$ reaction is usually
linked to its mirror companion, $^{3}H(^{3}H,2n)^{4}He$. A comparison of both
leads to a better understanding of the underlying dynamics and of the Coulomb
effects of reactions with three-cluster exit channels.

A first microscopic calculation for these reactions was presented in
\cite{kn:Vasi89}. A two-cluster approach was used for both the entrance and
exit channels. The nucleon-nucleon fragment cluster (denoted $NN$ for either
$pp$ or $nn$) carried a simple shell-model description, featuring a
pseudo-bound state with positive energy. The experimental cross-section or
$S$-factor at relatively high energy (approximately 1 MeV) was reproduced by
adjusting the Majorana exchange parameter of the effective $NN$-potential. The
available experimental data at the (small) energy range relevant in
astrophysical reactions were fairly well reproduced. In this model no
resonance state appears that would sufficiently amplify the $S$-factor in the
appropriate energy range, and thus would constitute an explanation for the
solar neutrino problem.%
\begin{figure}
[ptb]
\begin{center}
\includegraphics[
trim=0.000000in 0.206754in 0.000000in 1.039609in,
natheight=11.681000in,
natwidth=8.253800in,
height=7.6289cm,
width=10.6404cm
]%
{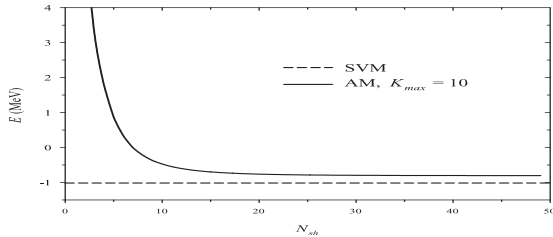}%
\caption{Ground state of $^{6}He$ as a function of the number of oscillator
shells $N_{sh}$ in the MJM three-cluster model, compared to the results of
\cite{Varga:1994sl}. The energy is relative to the $\alpha+n+n$ threshold.}%
\label{fig:figure11}%
\end{center}
\end{figure}

The previous model was further improved or enhanced in
\cite{kn:typel+91,kn:Desc94,kn:Csoto+Lang99} by using a more elaborate
description of the $NN$-channel, or by simulating the exit channel description
using both the $(^{4}He+N)+N$ and the $^{4}He+(N+N)$ two-cluster
configurations. The relative motion of the two clusters was described by a
discrete superposition of translated Gaussian functions. In each case the same
$S$-factor shape as a function of energy was obtained.

In this section we study both the $^{3}H\left(  ^{3}H,2n\right)  ^{4}He$ and
$^{3}He\left(  ^{3}He,2p\right)  ^{4}He$ reactions and use a correct treatment
of the corresponding three-cluster exit channels.

The bound state energy of $^{6}He$\ in the three-cluster description can
easily be obtained by diagonalizing the nuclear Hamiltonian. We have compared
it to the results obtained for the three-cluster calculation with the
Stochastic Variational Method (SVM) \cite{Varga:1995dm} \cite{Varga:1994sl}.
We have used the Minnesota potential without spin-orbit components, and an
oscillator parameter $b=1.285$\ fm which minimizes the $\alpha$-particle
energy as in \cite{Varga:1994sl}. In figure \ref{fig:figure11} we compare our
bound-state energy of $^{6}He$\ as a function of number of oscillator shells
$N_{sh}$\ to the SVM value. We used all HH's up to $K_{max}=10$. One notices
convergence for $N_{sh}\geq25$\ towards $E=-0.8038$\ MeV (relative to the
$\alpha+n+n\ $threshold). This is to be compared to $E=-1.016$\emph{\ }MeV for
the full SVM calculation; full convergence towards the SVM result would
require additional $K$ values, which is beyond the scope of the current
calculation.\emph{\ }These results encourage us to combine the two- and
three-cluster MJM descriptions to obtain an advanced description of the fusion
reactions $^{3}H(^{3}H,2n)^{4}He $ and $^{3}He(^{3}He,2p)^{4}He$.

\subsection{Model specifics}

We again rely on section \ref{sect:JMclusterModel}\ for the details of the
microscopic model. The specific cluster configurations used to describe the
six-nucleon systems $^{6}He$ and $^{6}Be$ are analogous to those of section
\ref{sect:ThreeClResonances}.

\subsubsection{A combined cluster model}

The six-nucleon wave functions will be built up by using both the two- and
three-cluster configurations, each one fully antisymmetrized:
\begin{equation}
\Psi_{L}=\mathcal{A}\left\{  \Psi_{3N}~\Psi_{3N}~f_{L}\left(  \mathbf{q}%
_{0}\right)  \right\}  +\mathcal{A}\left\{  \Psi_{\alpha}~\Psi_{N}~\Psi
_{N}~g_{L}\left(  \mathbf{q}_{1},\mathbf{q}_{2}\right)  \right\}
\label{eq:102}%
\end{equation}

The $\Psi_{A}$ ($N$ stands for either nucleon, $\alpha$ for $^{4}He$, and $3N$
for $^{3}H$ or $^{3}He$) represents cluster component wave functions, $f_{L}$
and $g_{L}$ refer to the wave functions of relative motion for the two-, and
three-cluster system. The $\mathbf{q}_{i}$ are Jacobi coordinates describing
the configuration of relative position of the clusters.

All of the reaction dynamics, and in particular the $S$-matrix elements, is
concentrated in the functions describing the relative motion, i.e. $f_{L}$ and
$g_{L}$ because the internal cluster wave functions are \textquotedblleft
frozen\textquotedblright.

We need to make a choice for the Jacobi coordinates $\mathbf{q}$ and for a
classification scheme of the wave-functions to be used in the expansion of
$f_{L}$ and $g_{L}$.

For the two-cluster configuration \cite{kn:Vasi89} (in $^{3}H+^{3}H$
respectively $^{3}He+^{3}He$) we use the standard spherical coordinates
$\mathbf{q}_{0}=\left\{  q_{0},\widehat{\mathbf{q}}_{0}\right\}  $ and take
the quantum numbers $\mu=\{n,L,M\}$ to classify the basis states. The $n$ is
the radial oscillator quantum number. As we have only central components in
the $NN$ interaction, the angular momentum $L$ of relative motion will be an
integral of motion for the system.

For the three-cluster configurations (in $^{4}He+p+p$ respectively
$^{4}He+n+n$) we use the Hyperspherical coordinates (\ref{eq:HyperCoords}).
This choice is consistent with the set of quantum numbers $\nu=\left\{
N,K,(l_{1}l_{2})LM\right\}  =\left\{  n_{\rho},K,(l_{1}l_{2})LM\right\}  $, in
which $N=2n_{\rho}+K$ represents the total number of oscillator quanta, and
$n_{\rho}$\ reflects the number of hyperradial excitations. The partial
angular momenta $l_{1}$ and $l_{2}$ are associated with the choice of Jacobi
vectors $\mathbf{q}_{1}$ and $\mathbf{q}_{2}$. A coupled $K$-channel
calculation with each channel characterized by the set of quantum numbers
$\nu_{0}=\left\{  K,(l_{1}l_{2})LM\right\}  $, has to be performed. This type
of basis is particularly suitable for the so-called Borromian nuclei, and
nuclei with pronounced three-cluster features, when the three-cluster
threshold represents the lowest energy decay channel.

\subsubsection{The boundary conditions}

The MJM boundary conditions are expressed in terms of the expansion
coefficients of the wave functions of relative motion. They are directly
connected to the boundary conditions in coordinate representation. For the
two-cluster configurations the asymptotic form of the expansion coefficients
in $f_{L}=\sum c_{n,L}\phi_{n,L}$ can be approximated by
\begin{equation}
c_{n,L}\simeq\sqrt{r_{n}}f_{L}\left(  r_{n}\right)  \label{eq:TwoClcn}%
\end{equation}
where the $\{\phi_{n,L}\}$ are the oscillator basis functions, and
$r_{n}=b\sqrt{4n+2L+3}$ is the classical turning point of the
three-dimensional oscillator with energy $E_{n}=\hbar\omega\left(
2n+L+3/2\right)  $. In three-cluster configurations the expansion is defined
by $g_{L}=\sum d_{n_{\rho},L}\phi_{n_{\rho},L}$, where the $\{\phi_{n_{\rho
},L}\}$ now stand for the oscillator basis for three-cluster relative motion.
The expansion coefficients behave asymptotically as
\begin{equation}
d_{n_{\rho},L}\simeq\sqrt{2}\rho_{n}^{2}g_{L}\left(  \rho_{n}\right)
\label{eq:ThreeCldn}%
\end{equation}
with $\rho_{n}=b\sqrt{4n_{\rho}+2K+6}$. For clarity we have indicated in the
preceding discussion only relevant indices to this section.

We consider both incoming and outgoing waves for the two-cluster
configurations
\begin{equation}
f_{L}\left(  \mathbf{q}_{0}\right)  \simeq\left[  \psi_{L}^{\left(  -\right)
}\left(  k_{0}q_{0}\right)  -S_{\{\mu\},\{\mu\}}~\psi_{L}^{\left(  +\right)
}\left(  k_{0}q_{0}\right)  \right]  Y_{LM}\left(  \widehat{\mathbf{q}}%
_{0}\right)
\end{equation}
where $S_{\{\mu\},\{\mu\}}$ is a notation to characterize the elastic
two-cluster scattering matrix element for the $^{3}H+^{3}H$, resp.
$^{3}He+^{3}He$ channels, and $Y_{LM}\left(  \widehat{\mathbf{q}}_{0}\right)
$ is the spherical harmonic.

Because we are only interested in reactions with a three-cluster exit-channel,
the asymptotic wave function can be written as
\begin{equation}
g_{L}\left(  \mathbf{q}_{1},\mathbf{q}_{2}\right)  =g_{L}\left(
\rho\mathbf{,}\Omega\right)  \simeq\sum_{\nu_{0}}\left[  -S_{\{\mu\},\{\nu
_{0}\}}~\psi_{K}^{\left(  +\right)  }\left(  k\rho\right)  \right]  H_{K}%
^{\nu_{0}}\left(  \Omega\right)
\end{equation}
where $S_{\{\mu\},\{\nu_{0}\}}$ is the scattering matrix element describing
the inelastic coupling between the two- and three-cluster channels, and
$H_{K}^{\nu_{0}}\left(  \Omega\right)  $ is the HH.

The total cross-section is given by the expression
\begin{equation}
\sigma\left(  E\right)  =\frac{\pi}{k_{0}^{2}}\sum_{L,S}\frac{\left(
2L+1\right)  \left(  2S+1\right)  }{4}\sum_{\nu_{0}}\left|  S_{\{\mu
\},\{\nu_{0}\}}\right|  ^{2}%
\end{equation}
with $S$ the total spin of the six-nucleon system.

As discussed in section \ref{sect:AsympSolCoordRep}, the asymptotic solutions
for incoming and outgoing waves can be written as
\begin{equation}
\psi_{\mathcal{L}}^{\left(  \pm\right)  }\left(  k\rho\right)  =\frac{1}%
{\sqrt{k}}W_{\pm i\eta,\lambda}\left(  \pm2ik\rho\right)  /\rho^{\frac
{\sigma-1}{2}} \label{eq:asymptSols}%
\end{equation}
where $W$\ are the Whittaker functions, and $\eta$\ the Sommerfeld
parameter\ (for the parameters $\mathcal{L}$, $\lambda$, $\sigma$\ and $\eta$
for two- and three-cluster channels: see table \ref{tab:HH state}).

By using the correspondences (\ref{eq:TwoClcn}) and (\ref{eq:ThreeCldn}) we
can now define the boundary conditions for the expansion coefficients
\begin{align}
c_{n,L}  &  \simeq\sqrt{r_{n}}\left[  \psi_{L}^{\left(  -\right)  }\left(
k_{0}r_{n}\right)  -S_{\{\mu\},\{\mu\}}~\psi_{L}^{\left(  +\right)  }\left(
k_{0}r_{n}\right)  \right] \nonumber\\
d_{n_{\rho},\nu_{0}}  &  \simeq\rho_{n}^{2}\left[  -S_{\{\mu\},\{\nu_{0}%
\}}~\psi_{K}^{\left(  +\right)  }\left(  k\rho_{n}\right)  \right]
\end{align}
or equivalently
\begin{align}
c_{n,L}  &  \simeq c_{n,L}^{\left(  -\right)  }-S_{\{\mu\},\{\mu\}}%
~c_{n,L}^{\left(  +\right)  }\nonumber\\
d_{n_{\rho},\nu_{0}}  &  \simeq-S_{\{\mu\},\{\nu_{0}\}}~c_{n_{\rho},\nu_{0}%
}^{\left(  +\right)  }%
\end{align}
using the notations
\begin{align}
c_{n,L}^{\left(  \pm\right)  }  &  \simeq\sqrt{r_{n}}\psi_{L}^{\left(
\pm\right)  }\left(  k_{0}r_{n}\right) \nonumber\\
d_{n_{\rho},\nu_{0}}^{\left(  \pm\right)  }  &  \simeq\rho_{n}^{2}~\psi
_{K}^{\left(  \pm\right)  }\left(  k\rho_{n}\right)
\end{align}

The matching of internal and asymptotic regions is equivalent to the one in
the traditional Resonating Group Method (RGM). The correspondence between the
matching point in coordinate space for RGM and in function space for MJM is
easily made (see \cite{kn:ITP+RUCA1}) through the value of the classical
oscillator turning point $r_{n}=b\sqrt{4n+2L+3}$\ for 2-cluster systems and
$\rho_{n}=b\sqrt{4n_{\rho}+2K+6}$\ for 3-cluster systems. An appropriate value
for the matching point can be obtained by choosing sufficiently large values
for the total number of oscillator quanta $N=2n+L=2n_{\rho}+K$\ in the
internal region.

\subsubsection{Shape analysis}

The HH's can reveal information on the spatial distribution of clusters and of
the reaction dynamics.

These harmonics define a probability distribution in 5-dimensional coordinate
(momentum) space for fixed values of hyperradius:
\begin{equation}
dW_{\nu_{0}}^{5}\left(  ~\Omega\right)  =\left\vert Y_{\nu_{0}}\left(
~\Omega\right)  \right\vert ^{2}d\Omega,\quad dW_{\nu_{0}}^{5}\left(
~\Omega_{k}\right)  =\left\vert Y_{\nu_{0}}\left(  ~\Omega_{k}\right)
\right\vert ^{2}d\Omega_{k}%
\end{equation}
By analyzing the probability distribution, one can retrieve the most probable
shape of three-cluster shape or \textquotedblleft triangle\textquotedblright%
\ of clusters. A full analysis of a function of 5 variables is non-trivial and
one usually restricts oneself to some specific variable(s). We integrate the
probability distribution $dW_{\nu_{0}}^{5}\left(  ~\Omega\right)  $ over the
unit vectors $\widehat{\mathbf{q}}_{1},\widehat{\mathbf{q}}_{2}$ (resp.
$\widehat{\mathbf{k}}_{1},\widehat{\mathbf{k}}_{2}$)
\begin{align}
dW_{\nu_{0}}\left(  \theta\right)   &  =\int\left\vert Y_{\nu_{0}}\left(
~\Omega\right)  \right\vert ^{2}\cos^{2}\theta\sin^{2}\theta d\theta
~d\widehat{\mathbf{q}}_{1}d\widehat{\mathbf{q}}_{2}\nonumber\\
dW_{\nu_{0}}\left(  \theta_{k}\right)   &  =\int\left\vert Y_{\nu_{0}}\left(
~\Omega_{k}\right)  \right\vert ^{2}\cos^{2}\theta_{k}\sin^{2}\theta
_{k}d\theta_{k}~d\widehat{\mathbf{k}}_{1}d\widehat{\mathbf{k}}_{2}%
\end{align}
and introduce the (new) variable(s)
\[
\mathcal{E}=\frac{q_{1}^{2}}{\rho^{2}}=\cos^{2}\theta,\quad\mathcal{E}%
=\frac{k_{1}^{2}}{k^{2}}=\cos^{2}\theta_{k}%
\]
In coordinate space these can be interpreted as the squared distance between
the pair of clusters associated with coordinate $\mathbf{q}_{1}$, or, in
momentum space, the relative energy of that pair of clusters. We obtain
\begin{align}
W_{\nu_{0}}\left(  \mathcal{E}\right)   &  =\frac{dW_{\nu_{0}}\left(
\theta\right)  }{d\theta}=\left\vert N_{K}^{\left(  l_{1},l_{2}\right)  }%
\cos^{l_{1}}\theta\sin^{l_{2}}\theta P_{n}^{\left(  l_{2}+1/2,l_{1}%
+1/2\right)  }\left(  \cos2\theta\right)  \right\vert ^{2}\cos^{2}\theta
\sin^{2}\theta\nonumber\\
&  =\left\vert N_{K}^{\left(  l_{1},l_{2}\right)  }\left(  \mathcal{E}\right)
^{l_{1}/2}\left(  1-\mathcal{E}\right)  ^{l_{2}/2}P_{n}^{\left(
l_{2}+1/2,l_{1}+1/2\right)  }\left(  2\mathcal{E}-1\right)  \right\vert
^{2}\sqrt{\mathcal{E}\left(  1-\mathcal{E}\right)  }%
\end{align}
This function represents the probability distribution for relative distance
between the two clusters, resp. for the energy of relative motion of the two
clusters. The kinematical factor $\cos^{2}\theta\sin^{2}\theta$ was included
to make $W_{\nu_{0}}\left(  \mathcal{E}\right)  $\ proportional to the
differential cross section in momentum space, provided the exit channel is
described by the single HH $Y_{\nu_{0}}\left(  \Omega\right)  $.%

\begin{figure}
[ptb]
\begin{center}
\includegraphics[
trim=0.000000in 0.129659in 0.000000in 0.659977in,
natheight=11.681000in,
natwidth=8.253800in,
height=12.5054cm,
width=10.6273cm
]%
{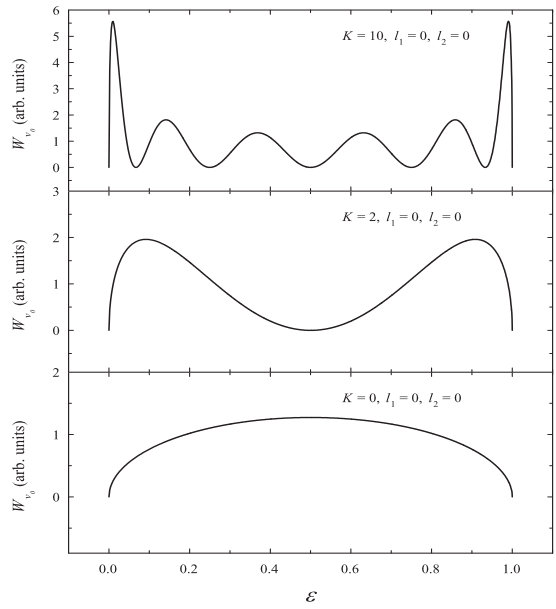}%
\caption{Function $W_{\nu_{0}}\left(  \mathcal{E}\right)  $ for $K=0,\ 2$ and
$10$ and $l_{1}=l_{2}=0$.}%
\label{fig:figure12}%
\end{center}
\end{figure}
In figure \ref{fig:figure12} we display\ $W_{\nu_{0}}\left(  \mathcal{E}%
\right)  $ for some HH's involved in our calculations. These figures show that
different HH's account for different shapes of the three-cluster systems. For
instance, the HH with $K=10$ and $l_{1}=l_{2}=0$ prefers the two clusters to
move with very small or very large relative energy, or, in coordinate space,
prefers them to be close to each other, or far apart.

\subsection{Results}

Again we use the VP as the $NN$ interaction. The Majorana exchange parameter
$m$ was set to be $0.54$\ which is comparable to the one used in
\cite{kn:Desc94}. The oscillator radius was set to $b=1.37$ fm (as in
\cite{kn:ITP+RUCA2,kn:Vasil97}) to optimize the ground state energy of the alpha-particle.

The VP does not contain spin-orbital or tensor components so that total
angular momentum $L$ and total spin $S$ are good quantum numbers. Moreover,
due to the specific features of the potential, the binary channel is uncoupled
from the three-cluster channel when the total spin $S$ equals 1; this means
that odd parity states $L^{\pi}=1^{-},2^{-},\ldots$ will not contribute to the
reactions.%
\begin{table}[tbp] \centering
\caption{Number of Hyperspherical Harmonics for $L=0$.}%
\begin{tabular}
[c]{|c|c|c|c|c|c|c|c|c|c|c|c|c|}\hline
$N_{ch}$ & 1 & 2 & 3 & 4 & 5 & 6 & 7 & 8 & 9 & 10 & 11 & 12\\\hline
$K$ & 0 & 2 & 4 & 4 & 6 & 6 & 8 & 8 & 8 & 10 & 10 & 10\\\hline
$l_{1}=l_{2}$ & 0 & 0 & 0 & 2 & 0 & 2 & 0 & 2 & 4 & 0 & 2 & 4\\\hline
\end{tabular}
\label{Tab:HH0}%
\end{table}%

To describe the continuum of the three-cluster configurations we considered
all HH's with $K\leq K_{max}=10$. In Table \ref{Tab:HH0} we enumerate all
contributing $K$-channels for $L=0$. For each two- and three-cluster channel
we used the same number $n=n_{\rho}=N_{int}$ of basis functions to describe
the internal part of\ the wave function $\Psi_{L}$. $N_{int}$ then also
defines the matching point between the internal and asymptotic part of the
wave function. We used $N_{int}$ as a variational parameter and varied it
between 20 and 75, which corresponds to a variation in coordinate space of the
RGM matching radius approximately between 14 and 25 fm. This variation showed
only small changes in the $S$-matrix elements, of the order of one percent or
less, and do not influence any of the physical conclusions. We have then used
$N_{int}=25$ for the final calculations as a compromise between convergence
and computational effort. We also checked the impact of $N_{int}$ on the
unitarity conditions of the $S$-matrix, for instance the relation
\[
\left\vert S_{\{\mu\},\{\mu\}}\right\vert ^{2}+\sum_{\nu_{0}}\left\vert
S_{\{\mu\},\{\nu_{0}\}}\right\vert ^{2}~=1
\]
We have established that from $N_{int}=15$ on this unitarity requirement is
satisfied with a precision of one percent or better. In our calculations, with
$N_{int}=25$, unitarity was never a problem. It should be noted that our
results concerning the convergence for the three-cluster system with a
restricted basis of oscillator functions agree with those of Papp et al
\cite{Papp:1999gd}, where a different type of square-integrable functions was
used for three-cluster Coulombic systems.

\begin{figure}
[ptb]
\begin{center}
\includegraphics[
trim=0.000000in 0.200725in 0.000000in 1.012597in,
natheight=11.213700in,
natwidth=7.923600in,
height=7.5059cm,
width=10.669cm
]%
{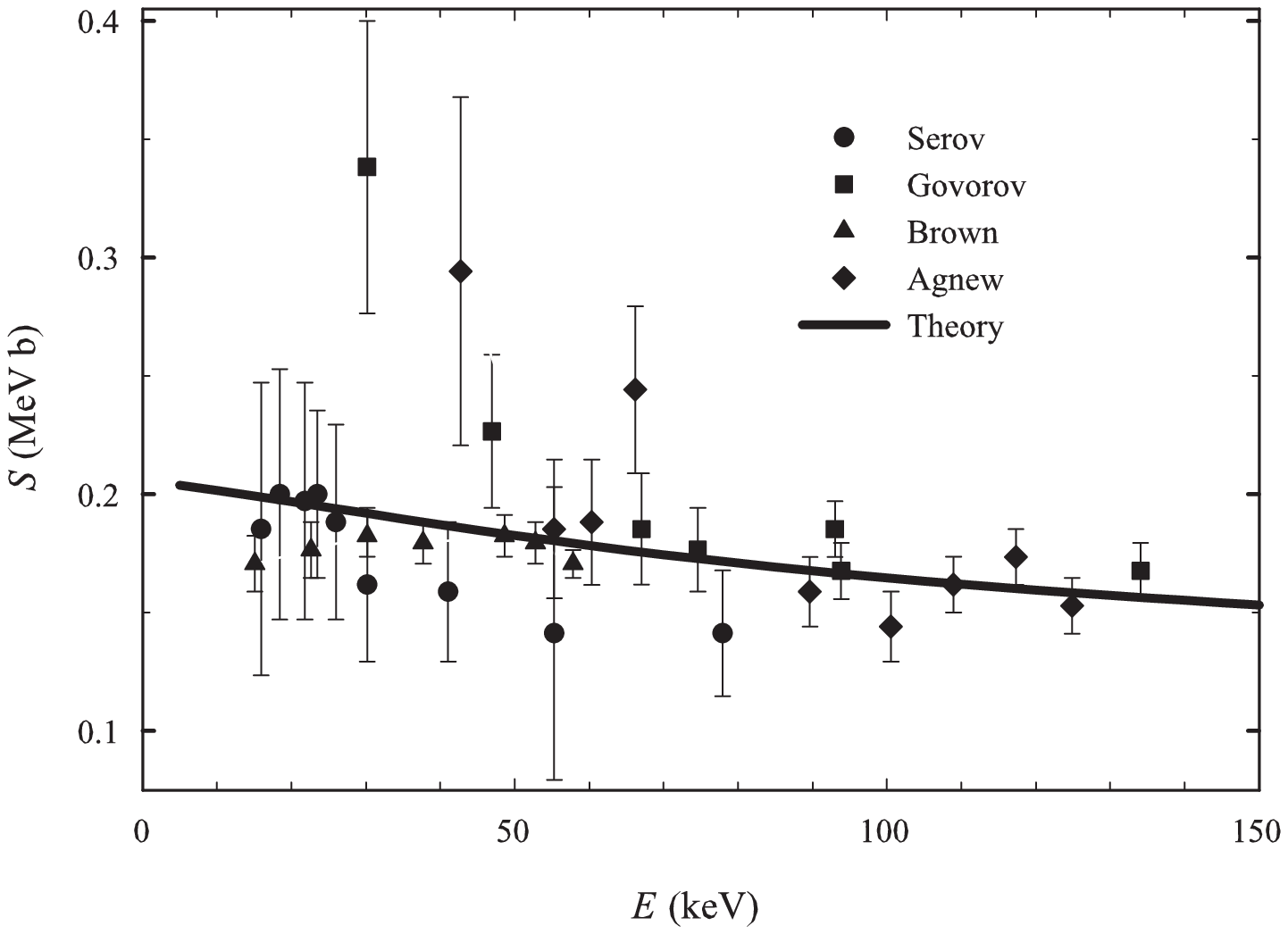}%
\caption{$S$-factor of the reaction $^{3}H\left(  ^{3}H,2n\right)  ^{4}He$.
Experimental data are taken from \cite{kn:serov77}(Serov), \cite{kn:govorov62}
(Govorov),\cite{kn:brown86} (Brown) and \cite{kn:agnew51} (Agnew)}%
\label{fig:figure13}%
\end{center}
\end{figure}

In figure \ref{fig:figure13} we show the total $S$-factor for the reaction
$^{3}H\left(  ^{3}H,2n\right)  ^{4}He$ in the energy range $0\leq E\leq200$
keV. One notices that the theoretical curve is very close to the experimental
data.%
\begin{figure}
[ptbptb]
\begin{center}
\includegraphics[
trim=0.000000in 0.200725in 0.000000in 1.012597in,
natheight=11.213700in,
natwidth=7.923600in,
height=7.5059cm,
width=10.5921cm
]%
{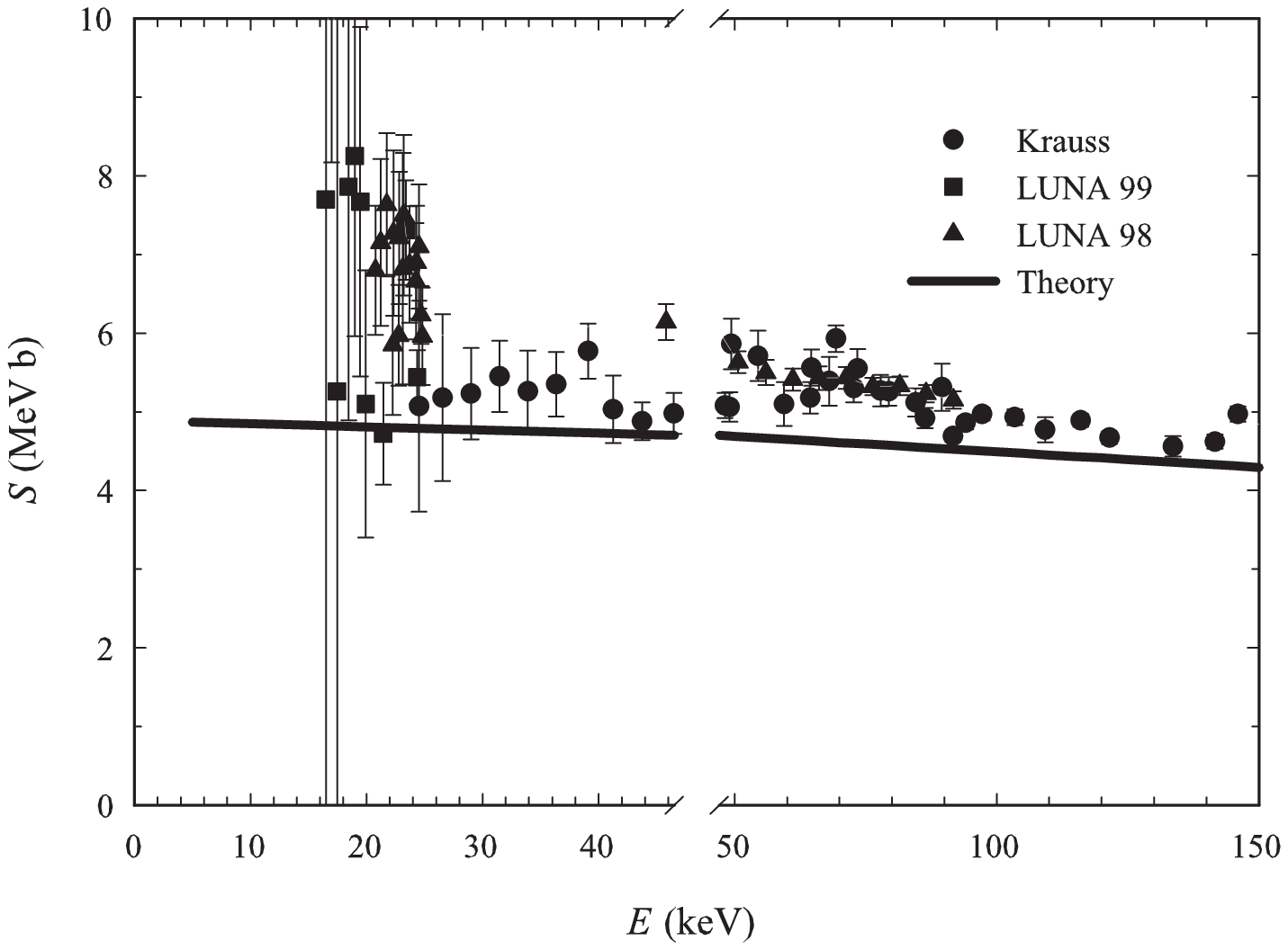}%
\caption{$S$-factor of the reaction $^{3}He\left(  ^{3}He,2p\right)  ^{4}He$.
Experimental data are from \cite{kn:krauss87} (Krauss), \cite{kn:LUNA99} (LUNA
99) and \cite{kn:LUNA98} (LUNA 98).}%
\label{fig:figure14}%
\end{center}
\end{figure}
The total $S$-factor for the reaction $^{3}He\left(  ^{3}He,2p\right)  ^{4}He$
is displayed in figure \ref{fig:figure14}. It is also close to the available
experimental data. The $S$-factor for both reactions is seen to be a monotonic
function of energy, and does not manifest any irregularities to be ascribed to
a hidden resonance. Thus no indications are found towards explaining the solar
neutrino problem.

The astrophysical $S$-factor at small energy is usually written as
\begin{equation}
S\left(  E\right)  =S_{0}+S_{0}^{\prime}E+S_{0}^{\prime\prime}E^{2}%
\end{equation}
We have fitted the calculated $S$-factor to this formula in the energy range
$0\leq E\leq200$ keV. For the reaction $^{3}H\left(  ^{3}H,2n\right)  ^{4}He$
we obtain the approximate expression:
\begin{equation}
S\left(  E\right)  =206.51-0.53~E+0.001~E^{2}\quad\text{keV b}%
\end{equation}
and for $^{3}He\left(  ^{3}He,2p\right)  ^{4}He$ we find:
\begin{equation}
S\left(  E\right)  =4.89-3.99~E+2.3~~10^{-4}~E^{2}\quad\text{MeV b}%
\end{equation}
One notices significant differences in the $S$-factor for the $^{6}He$ and
$^{6}Be$ systems. The $NN$-interaction induces the same coupling between the
clusters of entrance and exit channels for both $^{6}He$ and $^{6}Be$. It is
the Coulomb interaction that distinguishes both systems, and accounts for the
pronounced differences in the cross-sections and $S$-factors.%
\begin{figure}
[ptb]
\begin{center}
\includegraphics[
trim=0.000000in 0.196240in 0.000000in 0.989048in,
natheight=11.213700in,
natwidth=7.923600in,
height=7.7057cm,
width=10.7898cm
]%
{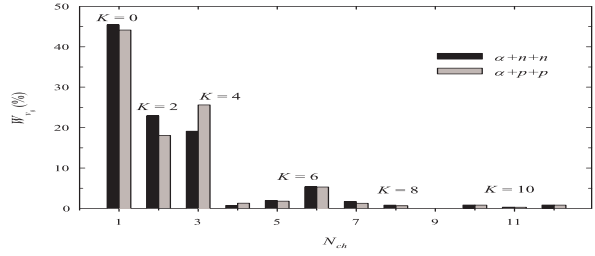}%
\caption{Three-cluster channel contributions to the total $S$-factor for the
reactions $^{3}H\left(  ^{3}H,2n\right)  ^{4}He $ and $^{3}He\left(
^{3}He,2p\right)  ^{4}He$ in a full calculation with $K_{max}=10$.}%
\label{fig:figure15}%
\end{center}
\end{figure}
\begin{figure}
[ptbptb]
\begin{center}
\includegraphics[
trim=0.000000in 0.198482in 0.000000in 1.003626in,
natheight=11.213700in,
natwidth=7.923600in,
height=7.5806cm,
width=10.8381cm
]%
{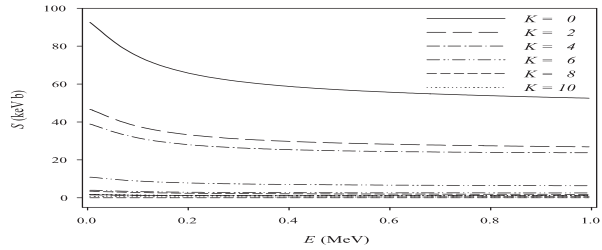}%
\caption{Three-cluster channel contributions to the total $S$-factor of the
reactions $^{3}H(^{3}H,2n)^{4}He$ in a full calculation with $K_{max}=10$, in
the energy range $0\leq E\leq1000$ keV.}%
\label{fig:figure16}%
\end{center}
\end{figure}

We compare the calculated $S$-factor to fits of experimental results for the
reaction $^{3}He\left(  ^{3}He,2p\right)  ^{4}He$:%

\begin{align}
S\left(  E\right)   &  =5.2-2.8~E+1.2~E^{2}\quad\text{MeV b
\cite{kn:dwarakanath71}}\nonumber\\
S\left(  E\right)   &  =\left(  5.40\pm0.05\right)  -\left(  4.1\pm0.5\right)
E+\left(  2.3\pm0.5\right)  E^{2}\quad\text{MeV b\cite{Arpesella:1997zf}%
}\nonumber\\
S\left(  E\right)   &  =\left(  5.32\pm0.08\right)  -\left(  3.7\pm0.6\right)
E+\left(  1.95\pm0.5\right)  E^{2}\quad\text{MeV b \cite{Bonetti:1999yt}}%
\end{align}
The constant and linear terms of the fit display a good agreement. The
difference in energy ranges between the calculated ($0\leq E\leq200$ keV) and
experimental ($0\leq E\leq1000$ keV) fits make it difficult to attribute any
significant interpretation to the discrepancy in the quadratic term.

The HH's method now allows to study some details of the dynamics of the
reactions considered.%
\begin{figure}
[ptb]
\begin{center}
\includegraphics[
trim=0.000000in 0.198482in 0.000000in 1.003626in,
natheight=11.213700in,
natwidth=7.923600in,
height=7.5806cm,
width=10.8381cm
]%
{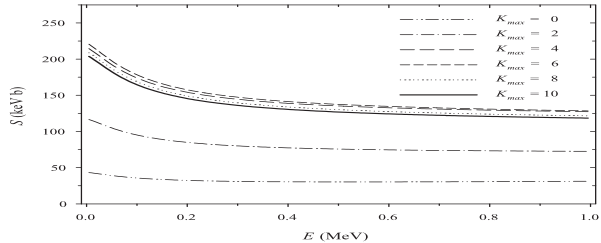}%
\caption{Convergence of the $S$-factor of the reaction $^{3}H(^{3}H,2n)^{4}He$
for $K_{max}$ ranging from $0$ to $10$.}%
\label{fig:figure17}%
\end{center}
\end{figure}
In Figs. \ref{fig:figure15} and \ref{fig:figure16} we show the different
three-cluster $K$-channel contributions ($W_{\nu_{0}}$) to the total
$S$-factor of the reactions. In figure \ref{fig:figure15} these contributions
(in \% with respect to the total $S$-factor) are displayed for some fixed
energy (1 keV), while figure \ref{fig:figure16} shows the dependency of
$W_{\nu_{0}}$ (in absolute value) on the energy of the entrance channel. One
notices that three HH's dominate the full result, namely the $\left\{
K=0;l_{1}=l_{2}=0\right\}  $, $\left\{  K=2;l_{1}=l_{2}=0\right\}  $ and
$\left\{  K=4;l_{1}=l_{2}=2\right\}  $, and this is true in both reactions.
The contribution of these states to the $S$-factor is more then 95 \% . There
also is a small difference between the reactions $^{3}H\left(  ^{3}%
H,2n\right)  ^{4}He$ and $^{3}He(^{3}He,2p)^{4}He$, which is completely due to
the Coulomb interaction.%
\begin{figure}
[ptbptb]
\begin{center}
\includegraphics[
trim=0.000000in 0.200725in 0.000000in 1.012597in,
natheight=11.213700in,
natwidth=7.923600in,
height=7.5037cm,
width=10.7415cm
]%
{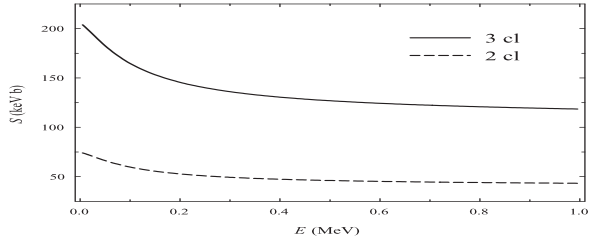}%
\caption{Comparison of the $S$-factor of the reaction $^{3}H(^{3}H,2n)^{4}He$
in a calculation with a three-cluster exit-channel and a pure two-cluster
model.}%
\label{fig:figure18}%
\end{center}
\end{figure}

The Figs. \ref{fig:figure15}\ and \ref{fig:figure16} yield an impression of
the convergence of the results. We notice that the contribution of the HH's
with $K>6$ is small compared to the dominant ones. This is corroborated in
figure \ref{fig:figure17} where we show the rate of convergence of the
$S$-factor in calculations with $K_{max}$ ranging from $0$ up to $10$. Our
full $K_{max}=10$ basis is seen to be sufficiently extensive to account for
the proper rearrangement of two-cluster configurations into a three-cluster
one, as the differences between results becomes increasingly smaller.

To emphasize the importance for a correct three-cluster exit-channel
description, we compare the present calculations\ to those in \cite{kn:Vasi89}%
, where only two-cluster configurations $^{4}He+2n$ resp. $^{4}He+2p$ were
used to model the exit channels. In both calculations we used the same
interaction and value for the oscillator radius. In figure \ref{fig:figure18}
we compare both results for $^{3}H(^{3}H,2n)^{4}He$. An analogous picture is
obtained for the reaction $^{3}He(^{3}He,2p)^{4}He$.%
\begin{figure}
[ptb]
\begin{center}
\includegraphics[
trim=0.000000in 0.109894in 0.000000in 0.569656in,
natheight=11.213700in,
natwidth=7.923600in,
height=14.0101cm,
width=10.7788cm
]%
{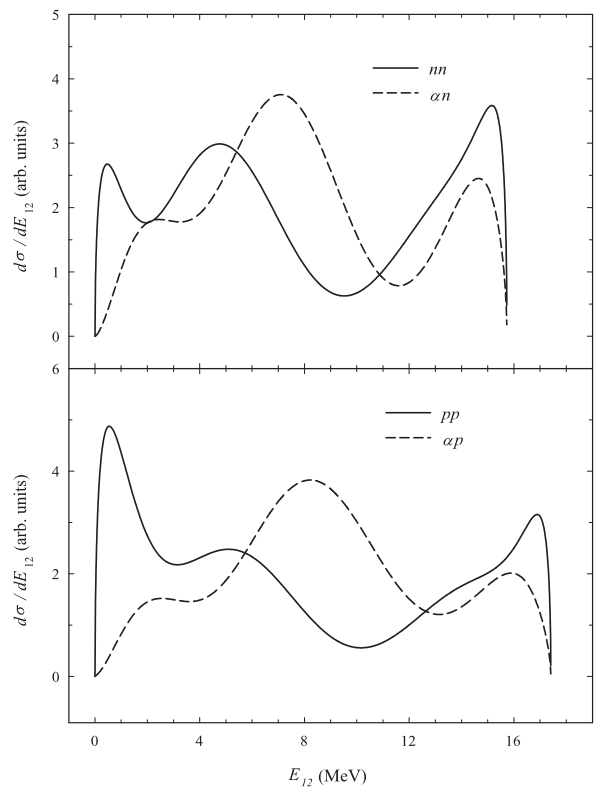}%
\caption{Partial differential cross sections of the reactions $^{3}%
H(^{3}H,2n)^{4}He$ and $^{3}He(^{3}He,2p)^{4}He$.}%
\label{fig:figure19}%
\end{center}
\end{figure}

\subsubsection{Cross sections}

Having calculated the $S$-matrix elements, we can now easily obtain the total
and differential cross sections. In this section we will calculate and analyze
one-fold differential cross sections, which define the probability for a
selected pair of clusters to be detected with a fixed energy $E_{12}$. To do
so we shall consider a specific choice of Jacobi coordinates in which the
first Jacobi vector $\mathbf{q}_{1}$ is connected to the distance between
these clusters, and the modulus of vector $\mathbf{k}_{1}$ is the square root
of relative energy $E_{12}$. With this definition\ of variables, the cross
section is
\begin{equation}
d\sigma\left(  E_{12}\right)  \sim\frac{1}{E}\int d\widehat{\mathbf{k}}%
_{1}d\widehat{\mathbf{k}}_{2}\left\vert \sum_{\nu_{0}}S_{\{\mu\}\{\nu_{0}%
\}}Y_{\nu_{0}}\left(  \Omega_{k}\right)  \right\vert ^{2}\sin^{2}\theta
_{k}\cos^{2}\theta_{k}d\theta_{k}%
\end{equation}%
\begin{figure}
[ptb]
\begin{center}
\includegraphics[
trim=0.000000in 0.158113in 0.000000in 0.795052in,
natheight=11.213700in,
natwidth=7.923600in,
height=9.7882cm,
width=10.043cm
]%
{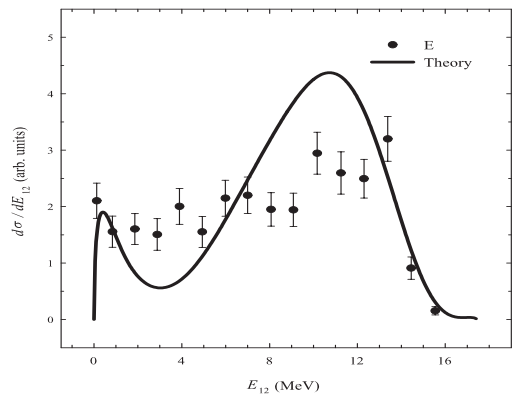}%
\caption{Calculated and experimental differential cross section for the
reaction $^{3}He(^{3}He,2p)^{4}He$. Experimental data is taken from
\cite{kn:dwarakanath71}. }%
\label{fig:figure20}%
\end{center}
\end{figure}
After integration over the unit vectors and substitution of $\sin\theta_{k}$,
$\cos\theta_{k}$, $d\theta_{k}$ with
\begin{align}
\cos\theta_{k}  &  =\sqrt{\frac{E_{12}}{E}};\quad\sin\theta_{k}=\sqrt
{\frac{E-E_{12}}{E}}\nonumber\\
d\theta_{k}  &  =\frac{1}{2}\frac{1}{\sqrt{\left(  E-E_{12}\right)  E_{12}}%
}dE_{12}%
\end{align}
one can easily obtains $d\sigma\left(  E_{12}\right)  /dE_{12}.$

In figure \ref{fig:figure19} we display the partial differential cross
sections of the reactions $^{3}H\left(  ^{3}H,2n\right)  ^{4}He$ and
$^{3}He\left(  ^{3}He,2p\right)  ^{4}He$ for the energy $E=10$ keV in the
entrance channel. The solid lines correspond to the case of two neutrons
(protons) with relative energy $E_{12}$, while the dashed lines represent the
cross sections of the $\alpha$-particle and one of the neutrons (protons) with
relative energy $E_{12}$.

We wish to emphasize the cross section in which two neutrons or two protons
are simultaneously detected. One notices a pronounced peak in the cross
section around $E_{12}\simeq0.5$ MeV. This peak is even more pronounced for
the reaction $^{3}He\left(  ^{3}He,2p\right)  ^{4}He$. It means that at such
energy two neutrons or two protons could be detected simultaneously with large
probability. We believe that this peak can explain the relative success of a
two-cluster description for the exit channels at that energy. The pseudo-bound
states of $nn$- or $pp$-subsystems used in this type of calculation then
allows for a reasonable approximation of the astrophysical $S$-factor.

Special attention should be paid to the energy range 1-3 MeV in the $^{4}He+n
$ and $^{4}He+p$ subsystems. This region includes $3/2^{-}$ and $1/2^{-}$
resonance states of these subsystems with the Volkov potential. In figure
\ref{fig:figure19} (dashed lines) we see that it yields a small contribution
to the cross sections of the reactions $^{3}H\left(  ^{3}H,2n\right)  ^{4}He$
and $^{3}He\left(  ^{3}He,2p\right)  ^{4}He$. This contradicts the conclusions
of \cite{kn:Desc94}\ and \cite{kn:Csoto+Lang99} where the $1/2^{-}$\ state of
the $^{4}He+N$ subsystem\ played a dominant role. We suspect this dominance to
be due to the interplay of two factors: the weak coupling between incoming and
outgoing channels, and the spin-orbit interaction.

In figure \ref{fig:figure20}\ we compare our results for the total proton
yield (reaction $^{3}He\left(  ^{3}He,2p\right)  ^{4}He$) to the experimental
data from \cite{kn:dwarakanath71}. The latter were obtained for incident
energy $E\left(  ^{3}He\right)  =0.19$ MeV. One notices a qualitative
agreement between the calculated and experimental data.

\begin{figure}
[ptb]
\begin{center}
\includegraphics[
trim=0.000000in 0.197361in 0.000000in 0.990170in,
natheight=11.213700in,
natwidth=7.923600in,
height=7.697cm,
width=12.2198cm
]%
{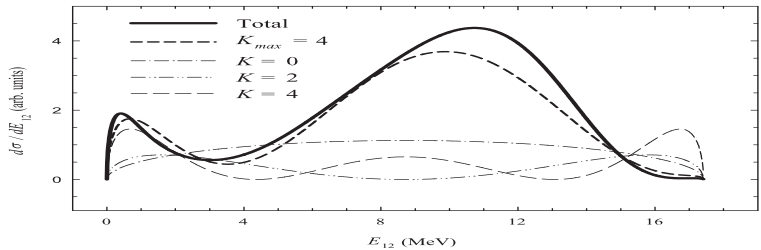}%
\caption{Partial cross sections for the reaction $^{3}He(^{3}He,2p)^{4}He$
obtained for individual $K=0,2$ and $4$ components, compared to the coupled
calculation with $K_{max}=4$ and the full calculations with $K_{max}=10$.}%
\label{fig:figure21}%
\end{center}
\end{figure}

The cross sections, displayed\ in Figs. \ref{fig:figure19} and
\ref{fig:figure20}, were obtained with the maximal number of HH's ($K\leq10$).
These figures should now be compared to the figure \ref{fig:figure12}, which
displays partial differential cross sections for a single $K$-channel. The
cross sections, displayed in Figs. \ref{fig:figure19} and \ref{fig:figure20},
differ considerably from those in Figs. \ref{fig:figure12} and comparable
ones, even for those HH's which dominate the wave functions of the exit
channel. An analysis of the cross section shows that the interference between
the most dominant HH's strongly influences the cross-section behavior. To
support this statement we display the proton cross sections obtained with
hypermomenta $K=0$, $K=2$, $K=4$ to those obtained with the full set of most
important components $K_{max}\leq4$ in figure \ref{fig:figure21}. One observes
a huge bump around 10 MeV which is entirely due the interference of the
different HH components. We also\ included the full calculation ($K_{max}%
\leq10$) to indicate the rate of convergence for this cross-section.%

\section{Conclusion}

In this paper we have presented a three-cluster description of light nuclei on
the basis of the Modified J-Matrix method (MJM). Key steps in the MJM
calculation of phase shifts and cross sections have been analyzed, in
particular the issue of convergence. Results have been reported for $^{6}He$
and $^{6}Be$. They compare favorably to available experimental data. We have
also reported results for coupled two- and three-cluster MJM calculations for
the $^{3}He\left(  ^{3}He,2p\right)  ^{4}He$ and $^{3}H\left(  ^{3}%
H,2n\right)  ^{4}He$ reactions with three-way disintegration. Again comparison
indicates good agreement with other calculations and with experimental data.

\end{document}